# Sensitive and accurate femtosecond pulse characterization via two-photon absorption in Fabry-Pérot laser diodes

Adrian F. Chlebowski*, Jakub Mnich, Lukasz A. Sterczewski, and Jaroslaw Z. Sotor**

*Laser & Fiber Electronics Group, Faculty of Electronics, Photonics and Microsystems, Wroclaw University of Science and Technology, Wybrzeze Wyspianskiego 27, 50-370 Wroclaw, Poland*

A B S T R A C T

Semiconductor lasers offer native bifunctionality enabling coherent light emission and linear photodetection. They can also operate as sensitive two-photon absorption detectors due to the third-order nonlinearity of the heterostructure constituting the active region. The strong two-photon response at room temperature is highly desired in ultrafast optics, where such detectors are used for interferometric characterization of femtosecond light pulses for shape and duration. Another niche is pulse detection in dual-comb ranging. While prior studies have focused on the two-photon response of commercial photodiodes or proprietary semiconductor microcavities for intensity autocorrelation measurements, a systematic analysis of the semiconductor lasers' ability to accurately characterize the optical pulse width is missing. To address this niche, here we measure autocorrelation traces of femtosecond pulses with varying durations from sub-55-fs to 260 fs at the common 1.5 µm wavelength using AlGaAs and InGaAsP laser diodes designed to operate at emission wavelengths of 0.95 µm and 1.3 µm, respectively. We validate the obtained waveforms using a silicon avalanche photodetector and conventional crystal-based second-harmonic autocorrelator. We consider the effects of optical polarization, operation mode and electrical load resistance on the shape and intensity of generated electrical signals. Our results prove the suitability of Fabry-Pérot laser structures for interferometric autocorrelation measurements of 53 µW (212 fJ pulse energy) average power pulses as short as 36 fs with a mean square error of $7 \times 10^{-3}$.

## 1. Introduction

Femtosecond laser pulses are indispensable in a variety of applications, including time-resolved spectroscopy [1], THz pulse generation [2], optical communications [3], and material processing [4], to name a few. Since direct optoelectronic detection at such timescales fails, sophisticated pulse characterization techniques are used instead to diagnose repetitive ultrafast events. Following the celebrated intensity autocorrelation (IAC) technique [5], various schemes incorporating a moving delay arm have been proposed to retrieve the pulse duration along with its temporal profile – a feature inaccessible by the basic implementation of IAC.

A prime example is field-resolved optical gating (FROG) [6], which measures a second-harmonic spectrum produced via the interaction of the characterized pulse with its time-delayed copy in a nonlinear crystal. A spectrogram image is recorded by varying the relative pulse delay, which next feeds an iterative phase recovery algorithm to reconstruct the femtosecond pulse waveform. Equally successful is the Spectral Phase Interferometry for Direct Electric Field Reconstruction (SPIDER) technique [7], which probes the interaction of two variable-delay unchirped pulse replicas with a strongly chirped and spectrally sheared (frequency-upconverted) pulse on a second-order nonlinear crystal. Unlike in FROG, no iterative numerical methods are needed to obtain the temporal pulse profile.

Although both of these techniques are well-suited for the accurate reconstruction of pulses even with single-fs durations, their practical implementation is typically complex and weakly compatible with low pulse energies (pJ or lower). This motivates the development of modalities of the experimentally simple IAC technique, like Phase and Intensity from Correlation and Spectrum Only (PICASO) [8] or Phase-Enabled Nonlinear Gating with Unbalanced Intensity (PENGUIN) [9]. These nonlinear autocorrelation techniques draw their strength from the experimental similarity to the linear Fourier transform spectroscopy technique employing a Michelson interferometer, where a pulse under test is split into two identical mutually delayed replicas. Next, depending on the detection scheme and interference capability, one can measure the ordinary linear spectrum (Fourier transformed *E*-field autocorrelation), or intensity autocorrelation (also with interference) if a nonlinear detection scheme is used.

The latter waveform is of interest here, whose measurement typically relies on second-order optical nonlinearity [10]. The optical pulse and its time-delayed copy are focused onto a nonlinear crystal with proper spatial orientation to provide phase and polarization matching for second harmonic generation (SHG). Only when there is sufficient temporal and spatial overlap between the pulses, the produced SHG signal can be detected using a sensitive photodiode (PD) or a photomultiplier tube (PMT) [11] as a function of time delay between the two pulse copies.

Unfortunately, traditional optical autocorrelators using optical crystals have several practical drawbacks. The nonlinear medium exhibits material dispersion, meaning that different wavelengths of light propagate through it at different group velocities [12]. For short pulses with broad spectral bandwidths, dispersion can lead to unintended pulse elongation and distortion of the autocorrelation trace. To minimize this effect, thin crystals are preferred at the expense of sensitivity. Also, efficient SHG generation requires precise phase matching between the fundamental second harmonic frequency. This is achieved by properly orienting the crystal with a suitable thickness. The biggest challenge, however, relates to the low efficiency of SHG, which requires a sensitive and bulky PMT for low-light signal detection.

* Corresponding author
*E-mail:* adrian.chlebowski@pwr.edu.pl, jaroslaw.sotor@pwr.edu.pl

Fortunately, many of these limitations can be circumvented by utilizing a third-order nonlinear process [13–18] – two-photon absorption (TPA) to produce a signal dependent on a product of optical intensities. Utilizing TPA semiconductors provides numerous advantages over traditional SHG methods. TPA – a third-order nonlinear process dependent on photon energies does not require phase matching [18–20] or beam polarization control [18,21], which greatly simplifies the experiment. With its high wavelength agility, TPA has already found application in spectral regions well beyond the near-infrared. Provided that the photon energy lies between half to the full energy bandgap of the photodetector material, wavelengths from the ultraviolet [22] to the infrared [17,23–25], can be covered. The detection process is also simplified since the TPA response and the conversion of the light signal into an electrical signal occur in a single semiconductor device. However, the most important advantage conveyed by TPA-based autocorrelators using semiconductor structures is its inherent sensitivity enabled by high nonlinear coefficients due to tight modal confinement in the waveguide [15, 17] accompanied by the natively high optical nonlinearity of the laser active region.

All these advantages have already been recognized by researchers. TPA-based autocorrelation measurements reported in the literature utilize various semiconductor structures like commercial photodiodes [14, 26–29], waveguides [15, 16, 30], PMTs [11], and light-emitting diodes [31,32] for temporal characterization of picosecond and femtosecond lasers. Some of the previous studies have also employed unbiased or reverse-biased edge-emitting laser diodes (LD) [18,26] as TPA detectors for picosecond pulses, due to their commercial availability, ease of integration, and waveguide-enhanced photocurrent. However, they were not systematically characterized in many respects in particular in different semiconductor material systems. It is uncertain how they disperse ultrafast (<50 fs) pulses and whether the photon lifetime inside the laser cavity (sub-mm long) has to be considered, or maybe absorption occurs over a small propagation distance inside the waveguide. Both effects impact the optical autocorrelation profile and hence the utility of lasers as practical detectors for pulse characterization.

Despite its popularity, measuring and *interpreting* the optical autocorrelation remains a challenging task and potentially leads to many controversies, as recently discussed by Fiehler et al. [33]. The most common is the claim of mode-locking and pulsed operation when a peak-to-background ratio of the interferometric autocorrelation deviates from the famous 8:1 ratio. This can easily arise if linear absorption is stronger than nonlinear one. Here, we adopt the community standards to avoid common errors in systematic studies on the response of commercial LDs to *femtosecond* pulses.

This study presents a systematic characterization of LDs based on popular semiconductor material systems for interferometric autocorrelation measurements in the near-infrared region. We validate the obtained traces using a silicon avalanche photodiode (Si APD), and a conventional crystal-based (CB) SHG autocorrelator. By examining pulse durations in the sub-55 fs to 260 fs range, we found a significant distortion of the autocorrelation trace for some LD devices. Additionally, we investigated the dependence the electrical signal on incident light polarization. In LDs, a variation in the polarization angle between the maximum and minimum signal positions results in a 27% reduction in peak-to-peak amplitude whereas in crystal-based systems, a similar change leads to an 82% decrease. Moreover, we explore how changes to operating point, including bias voltage and load resistance, influence the nonlinear response.

Sensitivity measurements reported here rely on the minimum detectable signal understood as a fringe-resolved autocorrelation (FRAC) trace with signal-to-noise ratio (SNR) of 2. Notably, LDs enable pulse width estimation at incident power levels as low as 53 µW, which remains beyond the reach of many conventional detection techniques.

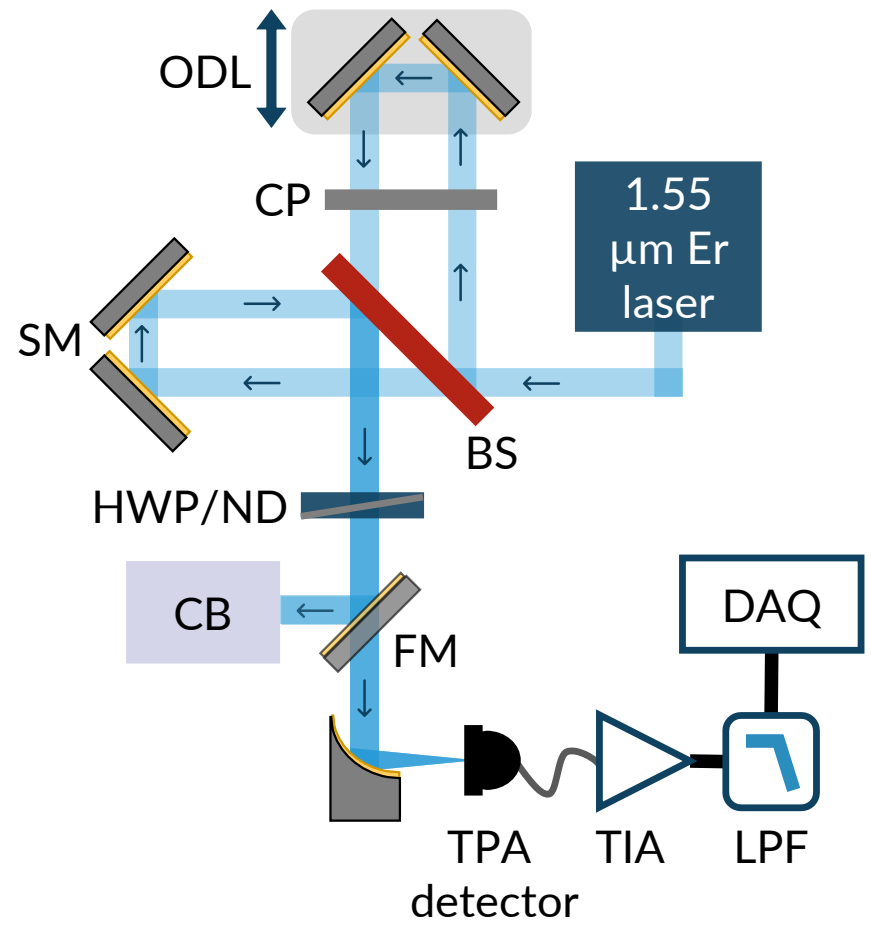

**Fig. 1.** Schematic diagram of the experimental setup. Optical pulses are generated by a mode-locked Er 1.5 µm fiber laser. They are next split and recombined utilizing a balanced Michelson interferometer, focused and detected by a TPA detector. Optical components constituting the setup: BS – beam splitter, CP – dispersion compensating plate, ODL – optical delay line, SM – stationary mirror, HWP – half-wave plate, ND – neutral density, FM – flip mirror, CB - crystal based SHG autocorrelator, TIA – transimpedance amplifier, LPF – lowpass filter, DAQ – data acquisition.

## 2. Experimental setup

To study the femtosecond pulse characterization capabilities of LDs, we measured interferometric autocorrelation traces using two Er-doped mode-locked fiber lasers with different repetition rates: 250 MHz, 90 MHz and average powers of 168 mW, 93.8 mW respectively at a central wavelength of 1550 nm. The first laser under test (LUT1) featured an adjustable pulse duration (sub-55 fs to 260 fs range) achieved through nonlinear amplification and spectral broadening, while the second (LUT2) exhibited a shorter pulse duration of sub-36 fs. The experimental setup consisted of a homemade interferometric autocorrelator built around a balanced Michelson interferometer. A voice-coil-type optical delay line (ODL) was used to actuate the moving arm and thus characterize pulses in a width range from 36 fs to 260 fs. For compatibility with the shortest pulse widths, an ultrafast broadband beamsplitter with controlled dispersion (ThorLabs UFBS50502) and an infrasil window (IW, ThorLabs UDP105) served for dispersion-managed interferometry. At the instrument output, a flip mirror was used to compare the obtained interferogram with a commercial device which was a CB autocorrelator (APE pulseCheck 50, crystal thickness 30 µm). To vary the illumination power reaching the photodetector, a neutral density filter (ND, ThorLabs NDK01) was positioned in front of the TPA photodetector itself.

Since the produced photocurrent lied in the nA–µA range, prior to digitization we conditioned the electrical signal from the LD by a low noise transimpedance amplifier (TIA, FEMTO DHPCA-100) and a third-order electrical low-pass filter (LPF, 10 kHz). Autocorrelation traces were recorded with a data acquisition device (DAQ, Oscilloscope Siglent SDS2354X) connected to a PC. Custom acquisition software was used to control the experiment. Due to imperfections in the motion of the scanning mirror, we digitally perform positioning correction using internally embedded quadrature encoder channels. The full width at half maximum (FWHM) of the autocorrelation trac is not a direct measure of the pulse duration. Instead, the pulse duration is estimated by fitting the autocorrelation trace with either $sech^2$ or Gaussian function and applying a deconvolution factor. For Gaussian shaped pulses, a deconvolution factor of 1.41 is used to extract the FWHM of the autocorrelation to the actual pulse duration. Since the LUTs are generated through nonlinear amplification and spectral broadening, the resulting pulses exhibit chirp, particularly near the centerburst of the autocorrelogram. As a result, the pulse duration is extracted by fitting a Gaussian function to the main lobe of the autocorrelation trace. The uncertainties in pulse width measurements were quantified by identifying the contributing sources of uncertainty and determining their standard deviations. The obtained values were propagated to simulation and mathematical analysis using Allan deviation and Monte Carlo methods. A detailed discussion of the uncertainty evaluation is provided in the supplementary material.

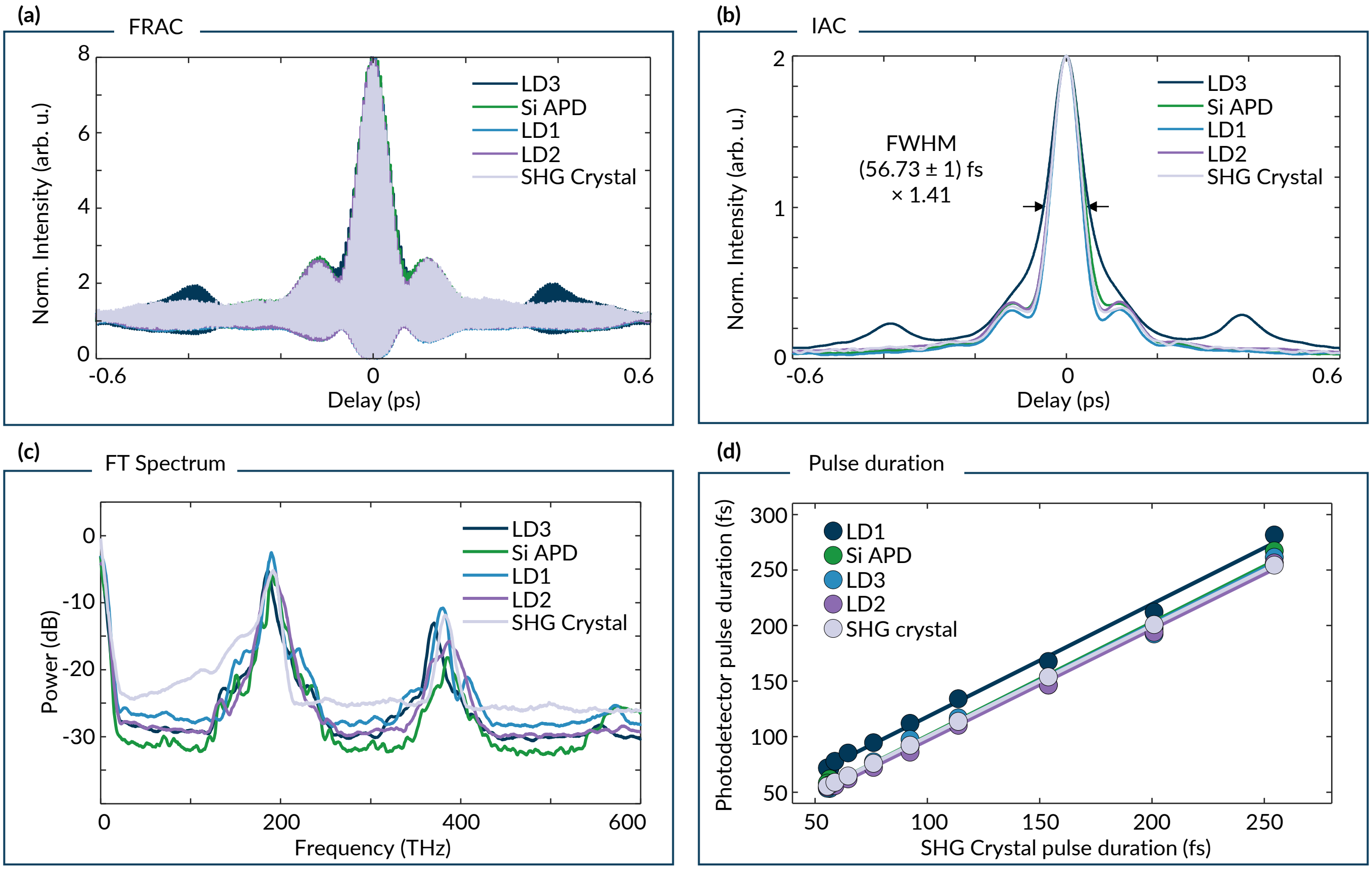

**Fig. 2** Two-photon photodetector comparison. (a) FRAC traces for a 1.5 µm wavelength laser with a deconvolved pulse width of (56.73 ± 1) fs. Traces obtained using the InGaAsP LDs, and the Si APD overlap with one probed using a SHG autocorrelator. Only the AlGaAs LD introduces distortions due to its pronounced dispersion. (b) IAC traces obtained via low-pass filtering of FRAC. (c) Power spectra of FRAC traces (d) Pulse duration of various detectors plotted against one measured using a thin-crystal SHG autocorrelator (ground truth).

## 3. Results

### 3.1 Material system comparison

The dimensions, composition, and refractive index profile of the waveguide strongly affect laser performance, including its output power, efficiency, temporal- and spectral properties. Moreover, laser media consisting of various materials exhibit diverse material bandgap energies. A condition for TPA to occur is that the photon energy be higher than half of the material bandgap energy, but cannot exceed it [19]. The AlGaAs material system yields TPA detectors operating from ~700 nm up to 1700 nm depending on the AlAs:GaAs ratio and hence the band-gap energy [34], whereas InGaAsP from 900 nm up to 3300 nm [34]. Simple Si-based PDs can also operate as two-photon detectors for 1550 nm with a spectral range from ~1200 nm up to 2200 nm [34].

In our experiment, using the same setup and electronics we compared three different commercial LDs serving as two-photon detectors: LD1 using the InGaAsP material system (NEC, NX7302AA-CC), average emitted power 0.2 mW and 1356 nm nominal wavelength emission, LD2 also based on InGaAsP (Luminent, MRLDFC010-NEB1), average power 1.2 mW and 1310 nm nominal wavelength emission, and LD3 based on AlGaAs (ABB, 1A137A-OC-2), average power 40 µW, 920 nm wavelength emission. In comparative tests, we also employed within the same setup a classic Si APD (Hamamatsu C12703-01) with the highest one-photon sensitivity at 800 nm. Ground truth autocorrelation traces were obtained using a traditional SHG thin-crystal-based autocorrelator. The light path was the same for all photodetectors to ensure that only the detector response was probed.

Figure 2a plots the results of the experiment, where FRAC traces obtained using the two InGaAsP LDs overlap with the CB SHG autocorrelation, thus proving the suitability of such devices for fs pulse characterization at telecom wavelengths. While minor deviations from the CB SHG autocorrelation can be observed in the case of the Si-based PD, the highest deviation occurs for the AlGaAs LD, whose dispersion Introduces FRAC distortions visible as pronounced wings away from the autocorrelogram centerburst. These discrepancies can be attributed to the waveguide geometry rather than bulk material dispersion. The InGaAsP diodes are designed to operate at longer wavelengths, which allows wider waveguides to be used for fundamental transverse (lateral) mode lasing. When the nominal emission wavelength is short (like for the AlGaAs laser at 920 nm), a narrow and hence more dispersive waveguide is typically used, which broadens and distorts the measured fs pulse. This motivates using LDs with nominal emission wavelengths closer to the probed one, at least from a waveguide geometry and related pulse dispersion perspective.

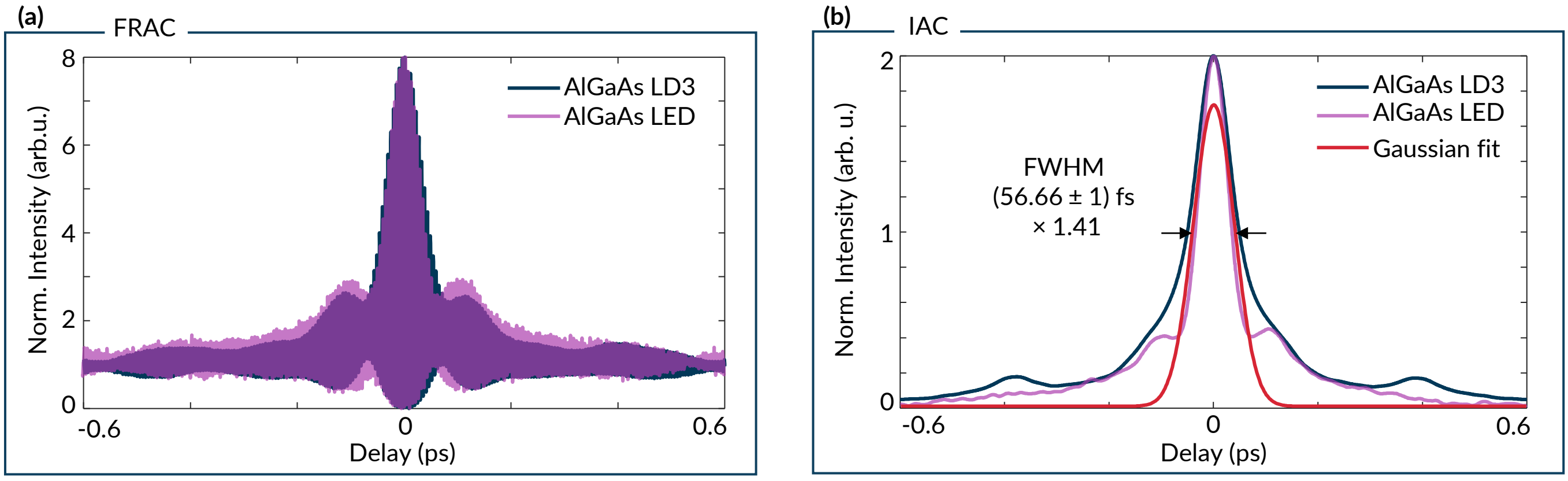

**Fig. 3.** Impact on the pulse distortion by waveguide geometry rather than the material system. (a) FRAC traces obtained using AlGaAs LD3 introduces distortions due to waveguide dispersion, in contrast to AlGaAs LED. (b) IAC traces obtained via low-pass filtering of FRAC with a LED deconvolved pulse width of (56.66 ± 1) fs.

The IAC (Fig. 2b) obtained via lowpass filtering of FRAC traces further supports the pulse distortion observation (Fig. 2a). Similar FWHMs (full width at half maximum) (Fig. 2b, 2d) are measured for LD1, LD2, and the Si APD, while the traces obtained using the AlGaAs LD deviates not only in width but also in autocorrelation shape. Spectral analysis (Fig. 2c), performed using Fourier transform (FT) indicates the presence of a second harmonic at a frequency of 392 THz, as expected for nonlinear detectors along with a weak (–25 dB) third-harmonic component, which is the most pronounced for the Si APD. A FWHM comparison (Fig. 2d) highlights the differences and the impact that the considered photodetectors have on pulse duration, which can be divided into two distinct groups. Group 1 includes non-distorted traces: InGaAsP LDs, Si APD, and the CB SHG autocorrelator. Group 2 includes the AlGaAs LD. From these measurements, one sees that the selection of a proper two-photon detector is critical to avoid introducing FRAC distortion and hence obtain an accurate pulse duration.

One of the semiconductor structures suitable for ultrashort pulse characterization is the light-emitting diode (LED), which due to the absence of a dispersive waveguide does not introduce broadening of pulse. To distinguish the effects of the waveguide geometry from those of the underlying material system on pulse elongation, next a comparative experiment was conducted between LD3 and a commercial LED fabricated from a analogous AlGaAs double heterostructure (VSMB3940X01, Vishay) emitting at 940 nm within the same setup and LUT.

The FRAC traces (Fig. 3a) and extracted pulse durations (Fig. 3b) from devices based on similar material platforms highlights the crucial influence of waveguide geometry. For pulses with a duration of (56.66 ± 1) fs, the LED-based autocorrelator preserves the shape of the autocorrelation trace, indicating minimal temporal distortion and reliable pulse duration. In contrast, LD3 exhibits significantly enhanced sensitivity which is increased 20 times. Specifically, LD1 and LD2—due to their favorable waveguide design—do not introduce noticeable degradation in pulse shape. Among them, LD1 demonstrated the most reliable performance and was therefore selected for further in-depth characterization.

### 3.2 Response to a varying average power and pulse duration

A semiconductor device capable of detecting TPA generates an electrical response proportional to quadratic signal intensity [19] when a pair of photons with energies in the range of half to near-full of the bandgap strikes its surface, preferably in a tightly focused manner. Considering the strong intensity dependence of this process, it is necessary to investigate the detector response depending on different excitation schemes [32]. The optical response was characterized by adjusting the excitation pulse width (from sub 55 to 260 fs) to investigate whether a LD acting as a two-photon detector introduces shape distortions and hence inaccurate pulse widths. Varying the pulse width was realized by adjusting the amplifier pump current. Therefore, the experimental arrangement was almost unchanged, with only the ND filters introduced to vary the average optical power delivered to the photodetectors (and hence the peak power).

LD1 detector was selected for comparative experiments. The average power was adjusted with three different ND filters, due to the reduced sensitivity of the SHG crystal, whereas for the LDs we used 5 filter grades. The average power values were selected so that the resulting signal did not significantly change the shape of autocorrelation traces, indicating that the resulting signal is not the minimum detectable signal. To assess the similarity between the different average powers, the envelope of each signal was extracted, and the mean squared error (MSE) was calculated. Low MSE values, such as $3.2\times10^{-3}$ (Fig. 4a), suggest a high degree of mutual agreement. The discrepancies observed between the SHG crystal and LD1 are primarily attributed to noise contributions. Moreover, with decreasing average power, and a decreasing pulse duration, a difference in sensitivity becomes noticeable. Fortunately, despite the strong dependence of TPA on peak intensity, varying the average optical power does not affect the autocorrelation shape (Fig. 4a-c), which maintains its desired peak-to-background ratio of 8:1 [33] for a mode-locked source.

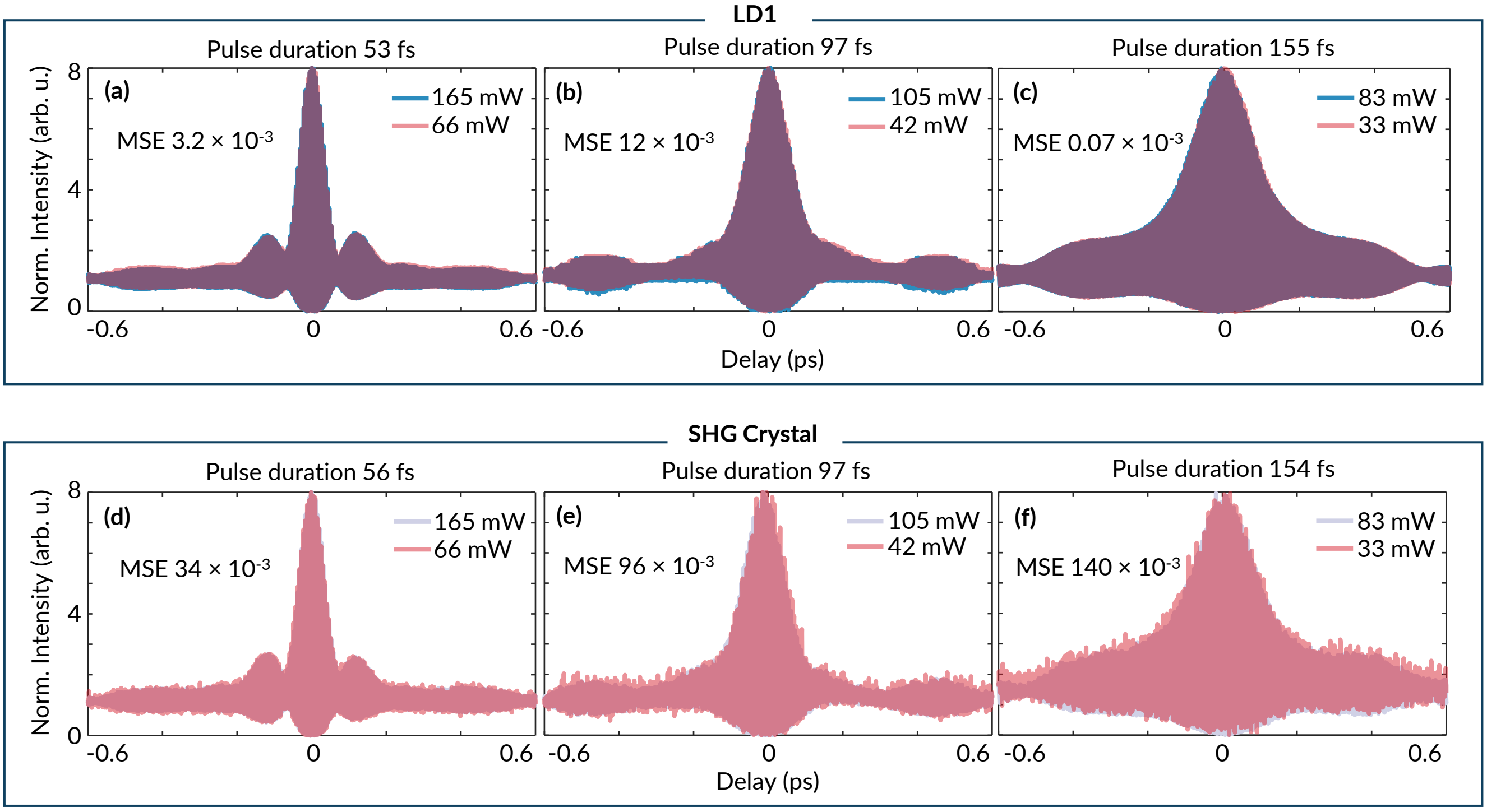

**Fig. 4.** Pulse durations obtained from the SHG crystal- and LD1-based autocorrelators at different levels of average excitation power. (a, d) Sub 53 fs pulse duration with 165 and 66 mW average power, (b, e) 97 fs with 105 and 42 mW, (c, f) 154 fs with 83 and 33 mW.

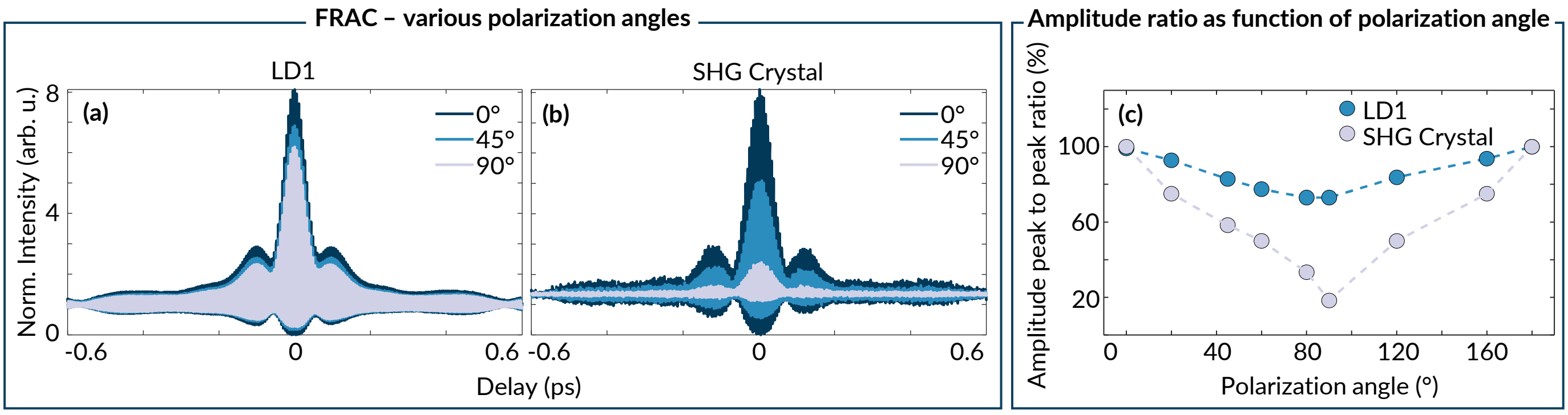

**Fig. 5.** Polarization dependence in (a) LD1 and (b) CB SHG autocorrelators. Varying the polarization angle had a significant impact on the peak-to-peak amplitude in the CB SHG autocorrelator, resulting in a signal reduction of up to 82% and noticeable distortion of the FRAC shape. In contrast, the LD1-based autocorrelator exhibited greater robustness, with a signal variation of ~27% between maximum and minimum peak-to-peak amplitudes. (c) Peak-to-peak amplitude ratio as a function of polarization angle.

### 3.3 Polarization dependence

Many correlation methods, including SHG and sum-frequency generation (SFG) in crystals, require careful polarization control for optimal performance [21]. In contrast, TPA lacks such restrictions, at least in principle. On the other hand, LDs, despite their high $\chi^{(3)}$ response, employ a sophisticated heterostructure constituting the quantum wells and a mode-selective waveguide, which naturally favors certain polarizations (quasi-TE or TM modes). Therefore, unlike in the case of a bulk Si PD, TPA LDs are expected to exhibit polarization sensitivity. To investigate this effect, the LD1's two-photon photocurrent was measured for different linear polarization orientations, which were varied using a half-wave plate (HWP, Thorlabs WPH10ME-1550). When illuminated by ~52.4 fs long pulses with 135 mW of average power, the peak-to-peak signal amplitude (converted from photocurrent to voltage using a TIA) varied from ~110 mV to 81 mV when the polarization is rotated about 90° with the strongest response for the quasi-TE-like mode (parallel to the LD's heterostructure layers). Modulating the polarization angle in LD1 does not alter the autocorrelation shape, preserving the desired peak-to-background ratio of 8:1 (Fig. 5a). A change in the polarization angle from 0° to 90° leads to a 27% reduction (Fig. 5c) in the maximum signal intensity, yet the autocorrelation shape remains intact, demonstrating the suitability of LD-based photodetectors in applications requiring polarization stability, such as fiber-optic systems. To explain the observed phenomena, we note that light absorption depends on the orientation of the electric field relative to the structure [35]. The atomic structure of GaAs and other semiconductors causes the TPA coefficient to change with polarization [36], which in turn affects the generation of electron-hole pairs and consequently the photocurrent. Also, waveguides support specific quasi-polarized optical modes (with a dominant TE/TM polarization component), resulting in different interaction forces with the detector material [37]. Similar measurements were performed using a CB SHG autocorrelator, naturally expected to exhibit significant signal changes to polarization angle due to the SHG phase matching requirement. First, we oriented the SHG crystal for maximum signal (referred to as 0° orientation), and next, we varied the state of polarization angle by rotating the waveplate. We found that a 45° change already lowers the FRAC signal by 40% (Fig. 5c).

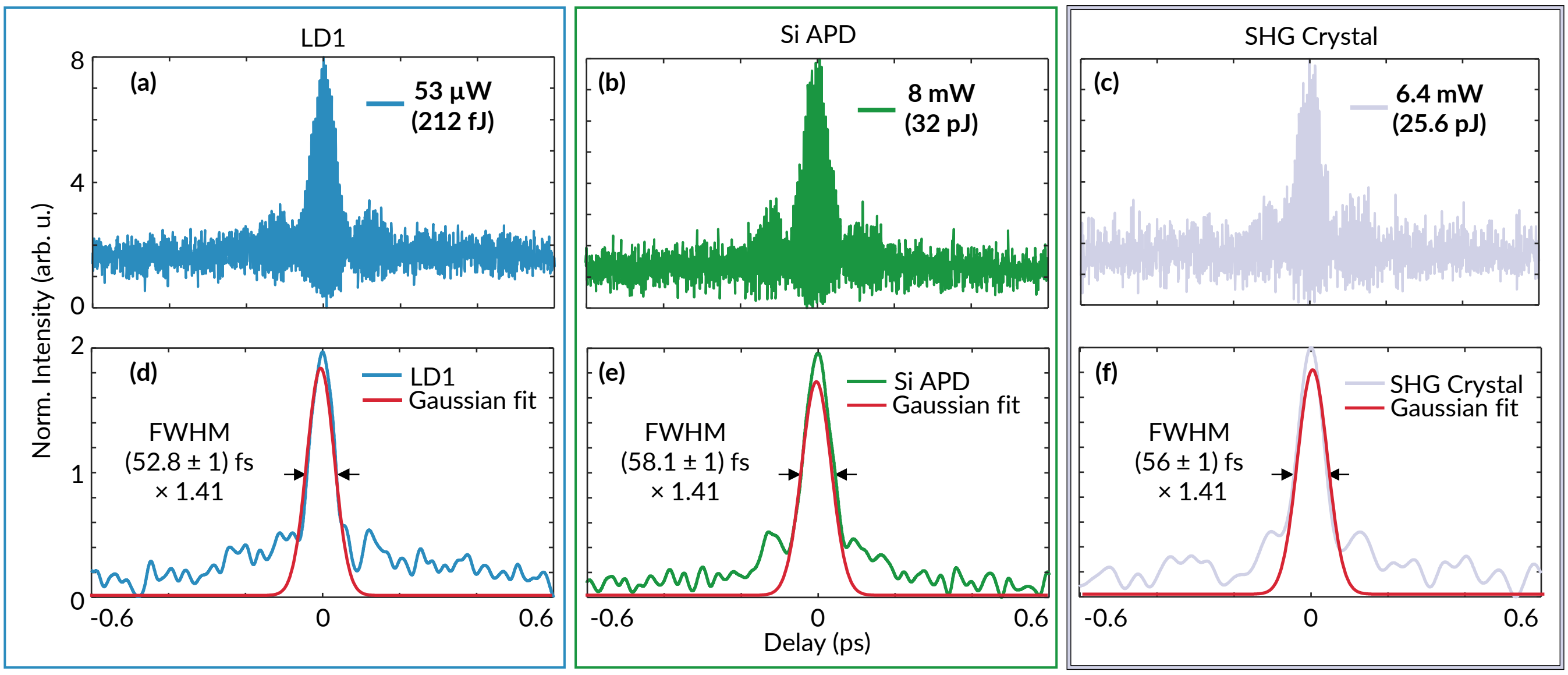

**Fig. 6.** Sensitivity comparison. (a, b) SHG Crystal, (c, d) Si APD, (e, f) InGaAsP LD. Despite low average optical powers, the obtained IAC accurately measures the pulse duration. The laser repetition rate was 250 MHz.

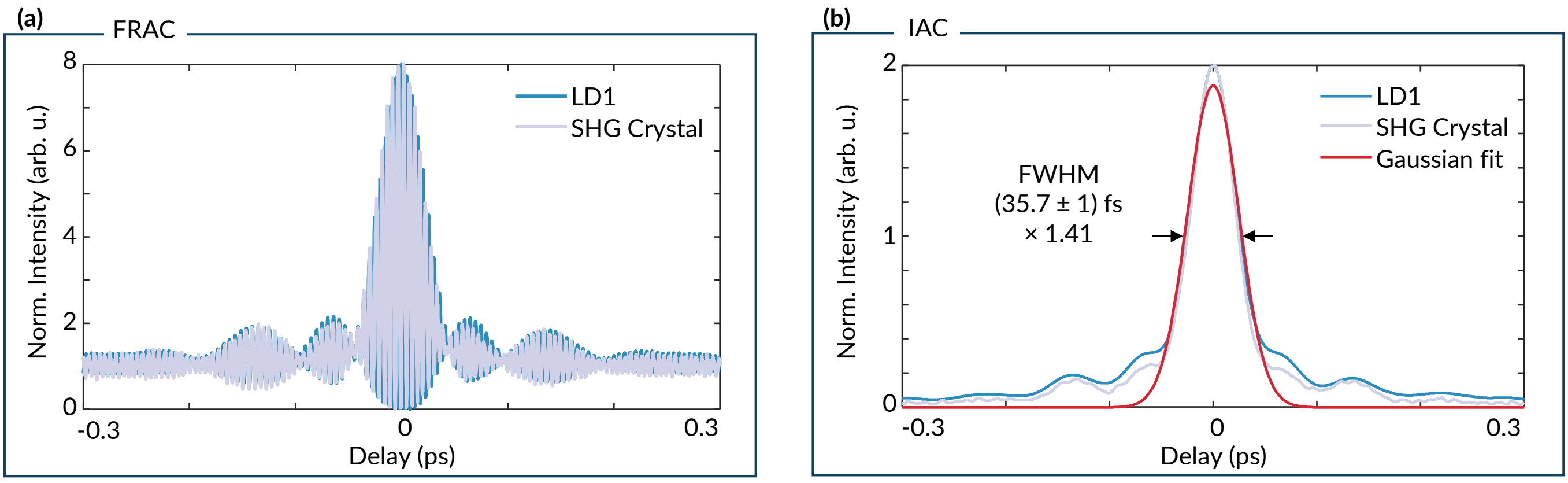

**Fig. 7.** FRAC (a) and IAC (b) with sub 36 fs pulse duration. Undistorted autocorrelation traces indicate faithful shape exactitude despite material dispersion. The laser repetition rate was 90 MHz.

Changing the angle of polarization by 90° causes the photodetector signal to drop by ~82% (Fig. 5c), distorting the shape of the autocorrelation and resulting in erroneous pulse duration measurements.

### 3.4 Sensitivity

In practice, nonlinear pulse characterization techniques require at least several pJ of pulse energy for multi shot pulse measurements and even more for single-shot measurements (if feasible at all [38]). Many experiments involving ultrafast nonlinear optical material characterization produce pulse energies in the fJ range or lower [38]. Fortunately, the sensitivity of semiconductor devices is significantly enhanced compared to their nonlinear crystal counterparts [18]. To assess this more quantitatively, the lowest detectable average powers were measured for LD1 Si APD, and a CB SHG autocorrelator: The lowest detectable signal was considered as one with an SNR of 2. The measurement did not use a lock-in amplifier, while the bandwidth was limited using a lowpass filter to 10 kHz. The lowest average power from the LD1-based autocorrelator was 53 µW (Fig. 6a, d, 212 fJ pulse energy), 8 mW (Fig. 6b, e, 32 pJ) for the Si APD, and 6.4 mW (Fig. 6c, f, 25.6 pJ) for one based on a SHG crystal. These values correspond to a sensitivity (defined as a product of peak and average intensities [18]) of $2{\cdot}10^2$ mW$^2$, $2.8{\cdot}10^6$ mW$^2$, and $4.1{\cdot}10^6$ mW$^2$, respectively. When normalized to the integration time (1 Hz bandwidth), one obtains values typical for high-bandwidth real-time pulse diagnostic systems [18] (here 2 mW$^2$, $5{\cdot}10^3$ mW$^2$, and $4.1{\cdot}10^4$ mW$^2$), with further improvements possible using a lock-in amplifier and narrowband detection away from the 1/f-noise rich spectral region.

Existing data indicates that the FP-diode based instrument offers a ~$10^3$ more sensitive detection for the given integration time compared to a CB SHG autocorrelator. One notable drawback with the Si APD is its responsiveness not only to 1.5 µm wavelength, but also to visible light. Residual ambient light distorts the second order autocorrelation, which makes this detection scheme somewhat impractical.

### 3.5 Shape exactitude

Autocorrelation employing SHG in nonlinear crystals, encounters challenges for pulses shorter than 10 fs due to material dispersion and phase mismatch [27]. Since such short pulses have broad spectra, pulse propagating at different group velocities in the nonlinear crystal lead to a temporal smearing and distortion of the autocorrelation [39]. In addition, the phase mismatch between the fundamental and the second harmonic waves limits SHG's performance for broad-spectrum pulses. To overcome the effects of dispersion and phase mismatch, very thin crystals (about 15 µm thick) [40] are applied; however, the production of these thin, polished crystals is expensive and complicated. In addition, the lowered thickness results in a lower conversion efficiency, requiring very sensitive PMTs to detect the signals. In order to investigate the shape exactitude in the short pulse regime, we employed LUT2, which generated the shortest available pulses in our setup, with a pulses duration of less than 33 fs. We found that the application of the LD1 introduces no additional distortion or artifacts to the measured autocorrelation function for a pulse width of sub 36 fs (Fig. 7b), thus supporting the accuracy in reproducing the pulse shape with a MSE $7\times10^{-3}$ (Fig. 7a). Based on these results, we infer that it's possible to accurately measure even shorter pulses without introducing distortion. However, overlap occurs not solely in the FRAC shape, but also in the IAC (Fig. 7b) and the FT spectrum. Autocorrelations using LD1 have less noise compared to traditional methods (Fig. 7a, 7b), while waveguide dispersion does not affect the shape of the FRAC. Distortions in the shape that are pronounced around the autocorrelogram centerburst are due to the presence of higher order dispersion caused by the LUT2 itself and occur in the case of both detectors employed. LD1 compared to GaAs microcavity devices has high quantum efficiency, which further enhances sensitivity [18]. Similar devices are widely implemented in fiber optic technology, which facilitates their integration into established systems.

### 3.6 Effect of operation mode on the shape and intensity of autocorrelation trace

Semiconductor devices for light detection, such as PDs and LDs, can operate in two distinct modes: photoconductive (biased) and photovoltaic (unbiased) [41]. The photoconductive mode is arguably more prevalent due to its numerous advantages. Applying a reverse bias voltage to the p-n junction expands the depletion region, which is beneficial in two primary ways. Firstly, the expanded depletion region enhances the photodetector sensitivity, making it more effective at converting light intensity into an electrical signal [42]. Secondly, a wider depletion region reduces the junction capacitance, resulting in a higher electrical bandwidth [43] and by allowing the photodetector to respond more rapidly to changes in light intensity. This mode ensures the linear (photocurrent vs light intensity) operation of the detector over a wide optical power range, and using a TIA keeps the detector's voltage nearly constant. In contrast, the photovoltaic mode transforms the photodetector into a small voltage source and has the advantage of virtually eliminating leakage current, thereby reducing dark current and noise levels, significantly improving the low-light performance [44]. However, the photovoltaic mode has notable drawbacks, including reduced linearity and dynamic range.

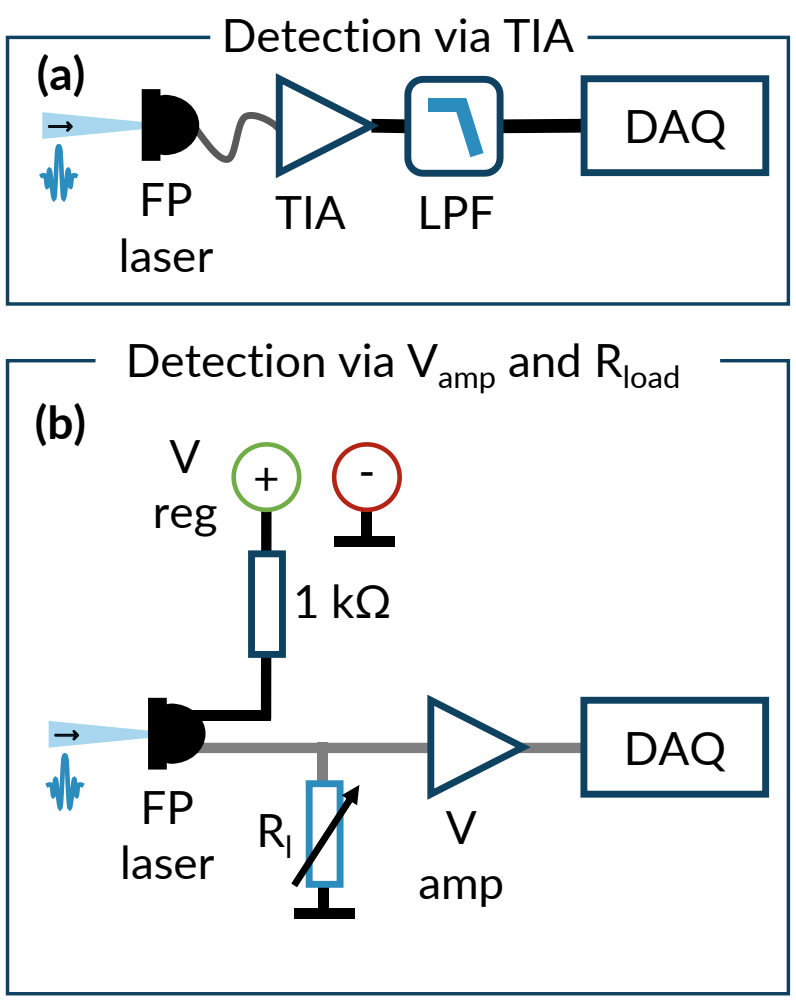

**Fig. 8.** (a) Detection via TIA. (b) Detection via voltage amplifier (V amp) and external load resistance ($R_l$). V reg – voltage regulator. V amp - voltage amplifier.

Additionally, the increased junction capacitance in this mode results in a reduced electrical signal bandwidth.

The above considerations are well-known and intuitive for simple semiconductor photodetectors. This intuition is lost in the case of complex heterostructure-based LDs operating in TPA mode, whose optical properties tune with electrical bias. Minor defects in the crystal structure may promote parasitic linear detection, particularly under some bias conditions. Therefore, the optimal mode of operation has to be empirically determined. For this purpose, we considered the biased and unbiased mode using a TIA (Fig. 8a), and without it when only an external load resistance was attached directly to the LD's electrodes (Fig. 8b). In all experiments, the electrical bandwidth was kept constant at 10 kHz (determined by the electrical LPF). No homodyne lock-in detection scheme was employed, but it should improve the lowest detectable signal further. Since the electrical signal generated during detection is typically in the nA range, a low noise voltage preamplifier (V amp, Stanford Research SR560) was employed for externally loaded LDs (i. e. not using a TIA). This amplifier also provided low-pass filtering to 10 kHz. The bias voltage was controlled via a programmable power supply (V reg, Rigol DP832), while the external load resistance ($R_l$, VT2, Thorlabs) was varied over a wide range from 50 Ω to 50 kΩ to assess its influence on the nonlinear response.

Figure 9 illustrates the influence of varying the bias voltage on the shape of the autocorrelation trace (Fig. 9a–d). Decreasing the bias voltage from unbiased to −0.6 V (Fig. 9b-c) resulted in no observable change in the peak-to-background ratio, which remained at the desired value of 8. This identifies −0.6 V as the lower limit beyond which the nonlinear response begins to degrade. As the bias voltage was further reduced, the peak-to-background ratio degraded, reaching 3.3 at −3.4 V (Fig. 9d). Applying a positive bias of +0.6 V also led to distortion, lowering the ratio to 6.2 (Fig. 9d). Similar behavior in the autocorrelation response was observed when comparing the use of an external load with a variable bias voltage to the internal biasing capabilities of the TIA. Specifically, for a bias voltage of −3.4 V (Fig. 9a), the peak-to-background ratio reached 3.1 with the external load configuration and 3.3 when adjusted internally via the TIA, indicating comparable levels of nonlinearity perturbation for both methods. Despite the reduction in peak-to-background ratio under bias, it is noteworthy that the peak-to-peak signal amplitude increased by 40% when comparing the unbiased condition to the -3.4 V bias.

Adjusting the load resistance for the unbiased case affected the linearity of the detection process (Fig. 9e–h). In the range of 50 Ω to 10 kΩ, the peak-to-background ratio remained constant at 8, indicating that the autocorrelation shape was not disturbed. A noticeable change appeared at 50 kΩ, where the ratio dropped to 3.6 (Fig. 9h).

Load resistances between 5 kΩ (Fig. 9f) and 10 kΩ (Fig. 9g) provided the most consistent results compared to the response obtained using the TIA (Fig. 9a–d). At bias voltages close to unbiased, small load resistances had little impact on the peak-to-background ratio, which remained 8 (Fig. 9e-g). However, as the negative bias increased, differences became more pronounced, at −3.4 V, a 50 Ω load obtained a ratio of 2.8, while a 5 kΩ load introduced 3.3. Increasing the load resistance also caused the peak-to-peak signal amplitude to rise linearly. The highest value was observed for 5 kΩ when comparing SNR across different load resistances. At −0.6 V bias (Fig. 9c), 5 kΩ reached an SNR of 29.2 dB, while 10 kΩ resulted in 27.7 dB. This shows that choosing an appropriate load resistance plays an important role in optimizing signal strength and nonlinear detection performance.

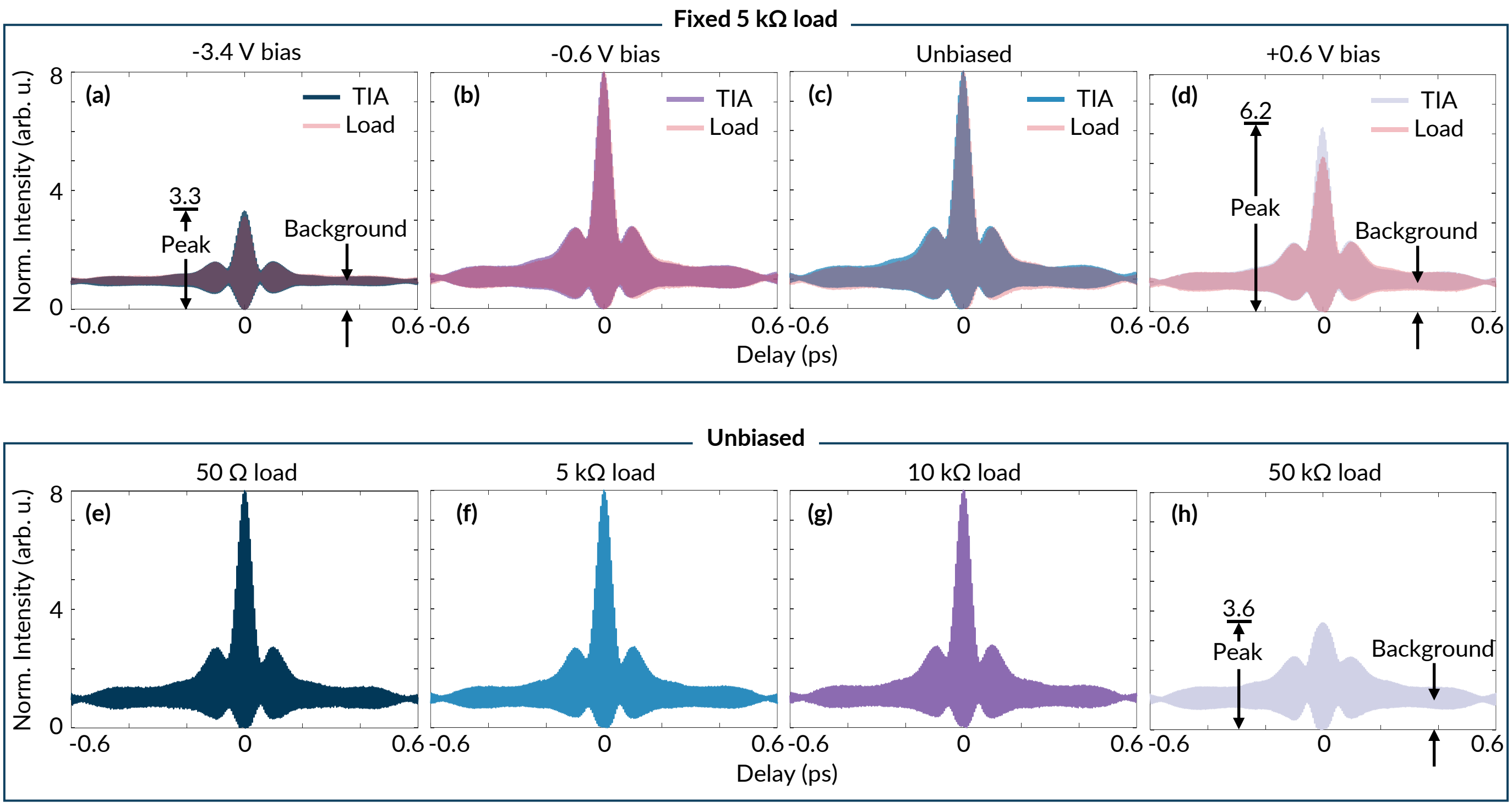

**Fig. 9.** Effect of (a, b, c, d) bias voltage and (e, f, g, h) load resistance on the shape of FRAC measured with LD1. Two configurations are considered: one using a transimpedance amplifier (TIA) keeping a constant voltage across the diode, and one with a variable load. Variations in bias voltage at a fixed load of 5 kΩ and TIA, with bias at -0.6 V and unbiased (b, c), do not affect the FRAC shape or peak-to-background ratio which remains 8. However, a bias of -3.4 V (d) and +0.6 V (a) disrupts linearity, introducing respectively a 3.3 and 6.2 ratio. The results converge when using both TIA and an external load. Adjusting the load resistance in unbiased mode did not affect the linearity of the detection response up to 10 kΩ (e-g). At 50 kΩ (h), the peak-to-background ratio reached 3.6.

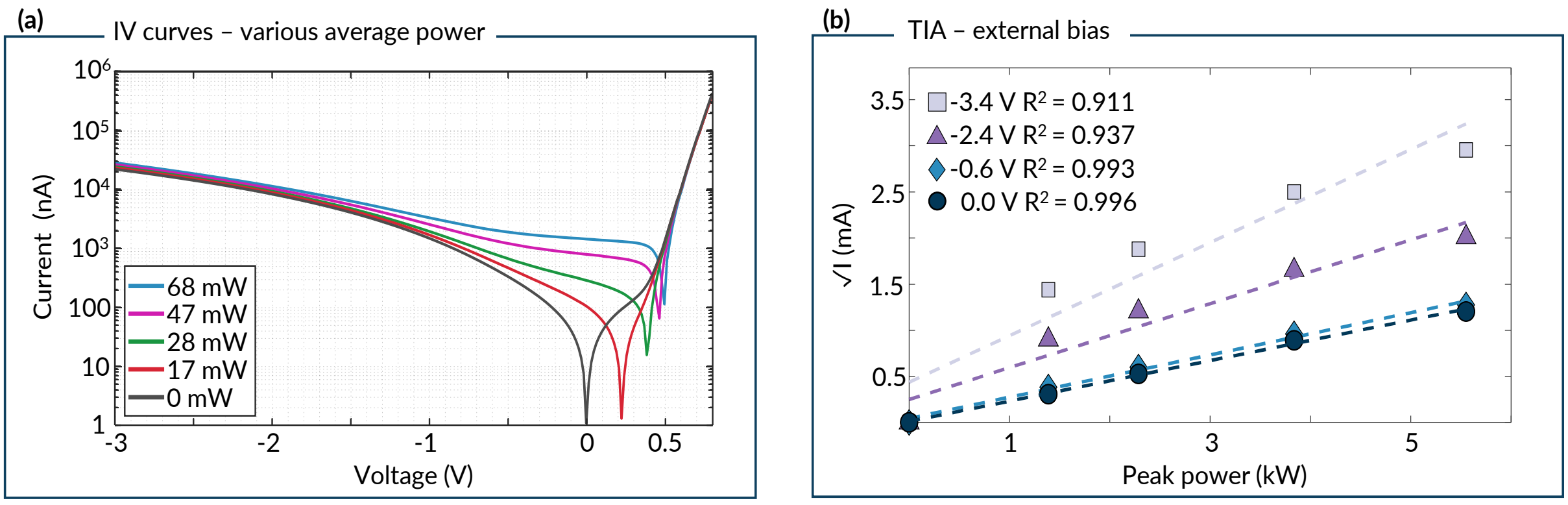

**Fig. 10.** (a) Log-scale of the dark and illuminated I-V response. (b) Response of the square root of the dark current compensated photocurrent depending on peak power via TIA at various bias voltages. Linear response occurs for 0 and -0.6 V, when $R^2$ is 0.996 and 0.993. Deviations from the linear response appear for -2.4 and -3.4 V, which reflects $R^2$ respectively 0.937 and 0.911.

### 3.7 Estimation the nonlinear response

In the previous section, we experimentally demonstrated that the choice of operating conditions—specifically bias voltage and load resistance—is critical for obtaining accurate autocorrelation measurements. While the experimental approach allows precise identification of optimal operating points, it requires the implementation of autocorrelator. To simplify the process especially for FP LDs, we propose an alternative method based on analyzing current-voltage (I-V) characteristics under varying illumination levels (Fig. 10a). By determining the corresponding load lines and extracting the responding photocurrent at its intersection (Fig. 10). Nominally, in the case of linear detectors, the photocurrent is directly characterized, as it scales linearly with the incident optical intensity. However, since the detector operates via TPA, its response is proportional to the square of the optical intensity. Despite this nonlinear behavior, the methodology remains analogous: by examining the element of the photocurrent that scales quadratically with peak power, it is still possible to assess the detector performance using standard approaches, with the understanding that the detected signal reflects a quadratic dependence on optical intensity. For complementary analysis, four different resistance values were selected to construct the load lines (intersecting with the LD's I-V curve): 50 Ω, 5 kΩ, 50 kΩ, and 2 MΩ (Fig. 11). The 50 Ω resistor represents a commonly used load standard; 5 kΩ provided a peak-to-background ratio comparable to TIA and yielded the highest SNR; 50 kΩ was selected due to its significantly reduced peak-to-background ratio (3.6) (Fig. 9h), indicative of nonlinearity in the detection process; 2 MΩ approximates the measured shunt resistance $R_{sh}$, with a steep load line slope allowing clear observation of I-V. For both 50 Ω and 5 kΩ, the square root of the photocurrent showed a linear dependence with a coefficient of determination ($R^2$) in the range of 0.997–0.993 on peak optical power at 0 and –0.6 bias voltages (Fig. 11b, c). As the bias increased, deviations from linearity emerged, consistent with the nonlinear response with $R^2$ in the range of 0.937 (Fig. 11c) observed in the corresponding autocorrelations (Fig. 9a). In contrast, the 2 MΩ resistance exhibited an inverse trend: the response was nonlinear at 0 and −0.6 V bias (Fig. 11a, b) with $R^2$ in range of 0.848–0.924, becoming more linear with $R^2$ of 0.998 as the bias voltage decreased (Fig. 11c).

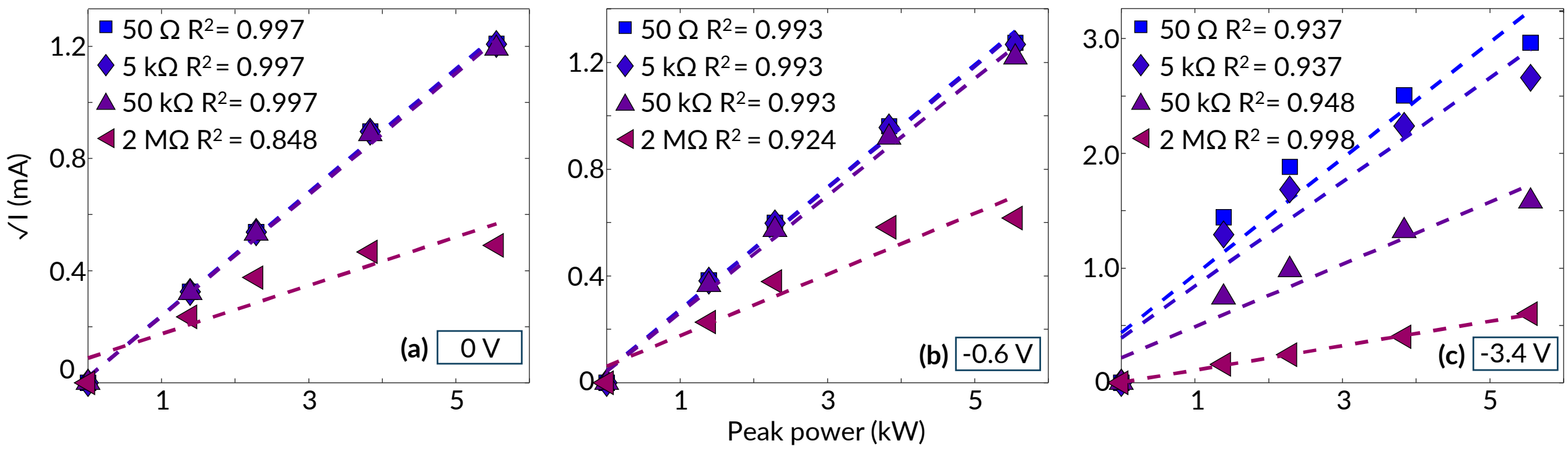

**Fig. 11.** Response of the square root of the photocurrent (compensated for the dark-current) depending on the peak optical power. Different external loads and different bias voltages are considered. At (a) unbiased and (b) −0.6 V voltages, resistances of 50 Ω–50 kΩ respond linearly with $R^2$ in the range of 0.993-0.997, and 2 MΩ nonlinearly with $R^2$ in the range of 0.924-0.848. At (c) -3.4 V, the response is nonlinear for resistances of 50 Ω-50 kΩ with $R^2$ in the range of 0.948-0.937 and linear for 2 MΩ with $R^2$ 0.998.

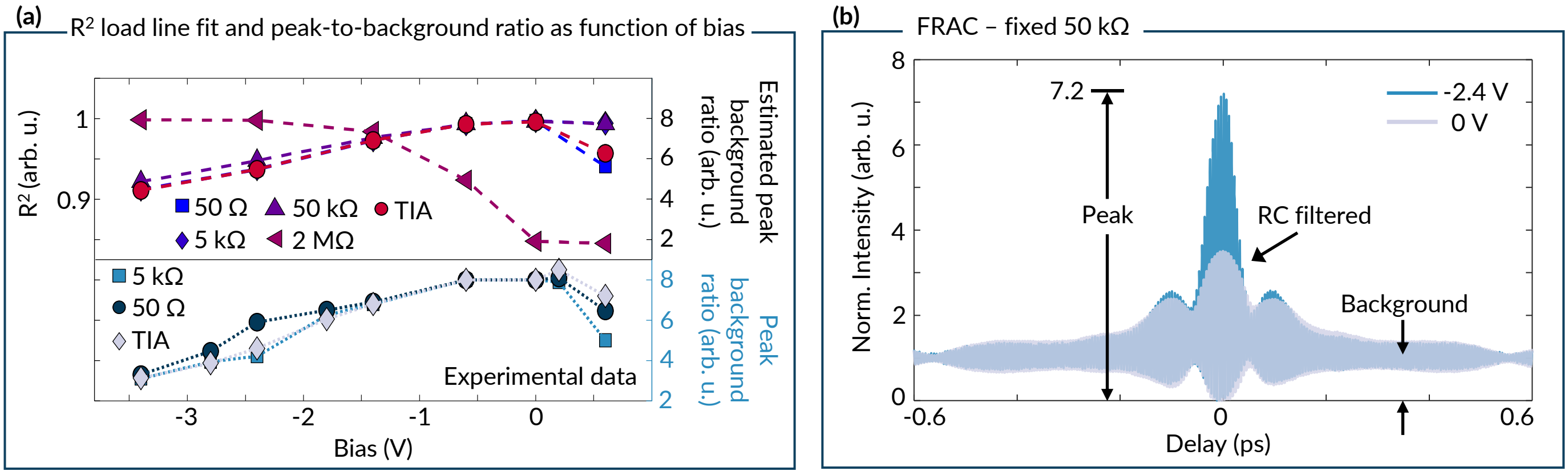

**Fig. 12.** (a) $R^2$ and peak-to-background ratio (FRAC 50 Ω, 5 kΩ, TIA) as function of bias voltage. (b) FRAC at 50 kΩ of load resistance for two biases. At 0 V bias, RC filtering degraded the signal to one with a peak-to-background ratio 3.6. For quasi-static measurements, the signal should have the desired 8:1 peak-to-background ratio. At −2.4 V, the increased detector speed due to the lowered junction capacitance yielded a peak-to-background ratio of 7.2.

$R^2$ derived from various bias voltages and load resistances can be used to identify optimal operating points where the peak-to-background ratio reaches the desired 8:1. Similar trends are observed when comparing peak-to-background ratios obtained from experimental autocorrelations with those predicted using $R^2$ (Fig. 12a). For voltages between 0 V and −0.6 V, the peak-to-background ratio remains at 8, consistent with experimental data. This correspondence also extends to −1.4 V, where both methods yield a ratio of 6.8. Deviations begin to appear as bias voltage is decreased, where at −2.4 V with a 50 kΩ load, the experimental peak-to-background ratio is 7.2 (Fig. 12b), while the predicted value is around 6 (Fig. 12a). At −3.4 V, the measured ratio drops to 3.3 (Fig. 9a, Fig. 12a), with the corresponding $R^2$-based estimate near 4.2. Interestingly, the 2 MΩ load exhibits an opposite trend. As the bias voltage decreases, the peak-to-background ratio increases, reaching 8 at −2.4 V. This behaviour suggests that, under specific conditions, high-impedance configurations may compensate for nonlinear distortions through voltage-dependent changes in the diode's internal properties, providing a useful strategy for tuning the detection response.

The observed discrepancies between experimental results and theoretical estimates (Fig. 12a) can be attributed to the distinction between static and dynamic measurement conditions. Static measurements—based on I-V curves (Fig. 10, 11, 12a)—reveal the inherent nonlinearity and electrical characteristics of the detector at fixed operating points, without the presence of a time-varying signal. These measurements are useful for assessing the detector's baseline behaviour but do not account for time-dependent effects.

In contrast, dynamic effects arise from the motion of the ODL during actual autocorrelation measurements (Fig. 9). A system that exhibits nonlinear behaviour under static conditions may not respond in the same way dynamically if the scan speed is too high. In such cases, the RC time constant—determined by the junction capacitance and load resistance—can introduce bandwidth limitations that distort the autocorrelation signal (Fig. 11b). This may lead to a degradation in the peak-to-background ratio, even when the static nonlinearity appears favourable. A clear example of this is observed with the 50 kΩ load. Although the I-V curves (Fig. 10) and calculated $R^2$ values indicate nonlinear response (Fig. 11a), the high capacitance and resistance result in significant signal averaging due to the RC effect (Fig. 12b), reducing the peak-to-background ratio. This limitation becomes evident when comparing results at 0 V and -2.4 V (Fig. 12b): the ratio changes from 3.6 to 7.2 simply by applying a negative bias, which reduces the junction capacitance and effectively increases the system bandwidth. This highlights the importance of considering both static and dynamic behaviours when determining optimal operating conditions for ultrafast pulse characterization.

## Conclusion

In summary, we systematically characterized FP LDs operating as two-photon absorption detectors. Three laser diodes based on AlGaAs and InGaAsP material systems were evaluated and compared against established standards, including a SHG crystal-based autocorrelator, silicon avalanche photodetector and a light-emitting diode. Using pulse durations ranging from 50 fs to 250 fs, we demonstrated that shorter-wavelength laser diodes (here based on the AlGaAs at 920 nm of nominal emission wavelength) may be unsuitable for ultrafast pulse characterization, as they introduce distortions in the autocorrelation trace and pulse duration attributable to their increased waveguide dispersion. The latter is dictated by the lower emission wavelength necessitating smaller waveguide dimensions for single-transverse mode operation. In contrast, the InGaAsP material system underlying two longer-wavelength FP diodes (~1300 nm) faithfully preserved both the autocorrelation shape and pulse duration, accurately measuring pulses as short as sub-36 fs with a MSE of $7\times10^{-3}$.

We also examined the impact of polarization angle on the detection process. The InGaAsP-based FP LD maintained a peak-to-background ratio of 8 across various angles of polarization, with photocurrent decreasing by only 27% relative to the maximum peak-to-peak amplitude. This is in stark contrast to conventional CB SHG autocorrelators, where a 90° change in polarization angle leads to distortion of the autocorrelation shape and a signal drop of up to 82%.

One of the key advantages of FP laser LDs as detectors is their strong TPA response at room temperature, offering a sensitivity enhancement of $\sim10^3$ compared to SHG CB autocorrelators. As a result, durations can be estimated at average powers as low as 53 μW - levels typically unreachable with conventional techniques.

In addition, we investigated both unbiased and biased operation modes and their influence on autocorrelation shape and peak-to-background ratio. Experimental results indicated that − 0.6 V is the lower bias limit beyond which the nonlinear response begins to degrade. Finally, by analyzing illuminated current-voltage characteristics, we identified trends that enable prediction of optimal operating points without the need to construct a full autocorrelator setup.

## Data availability

Data underlying the results presented in this paper are available in the dataset under Ref. [45]

## References

[1] M. Maiuri, M. Garavelli, G. Cerullo, Ultrafast Spectroscopy: State of the Art and Open Challenges, J. Am. Chem. Soc. 142 (2020) 3–15. https://doi.org/10.1021/jacs.9b10533.

[2] L. Zhang, K. Mu, Y. Zhou, H. Wang, C. Zhang, X.-C. Zhang, High-power THz to IR emission by femtosecond laser irradiation of random 2D metallic nanostructures, Sci. Rep. 5 (2015) 12536. https://doi.org/10.1038/srep12536.

[3] V. Torres-Company, J. Schroder, A. Fulop, M. Mazur, L. Lundberg, O.B. Helgason, M. Karlsson, P.A. Andrekson, Laser Frequency Combs for Coherent Optical Communications, J. Light. Technol. 37 (2019) 1663–1670. https://doi.org/10.1109/JLT.2019.2894170.

[4] Y. Jia, F. Chen, Recent progress on femtosecond laser micro-/nano-fabrication of functional photonic structures in dielectric crystals: A brief review and perspective, APL Photonics 8 (2023) 090901. https://doi.org/10.1063/5.0160067.

[5] H.P. Weber, Method for Pulsewidth Measurement of Ultrashort Light Pulses Generated by Phase-Locked Lasers using Nonlinear Optics, J. Appl. Phys. 38 (1967) 2231–2234. https://doi.org/10.1063/1.1709859.

[6] D.J. Kane, R. Trebino, Characterization of arbitrary femtosecond pulses using frequency-resolved optical gating, IEEE J. Quantum Electron. 29 (1993) 571–579. https://doi.org/10.1109/3.199311.

[7] C. Iaconis, I.A. Walmsley, Spectral phase interferometry for direct electric-field reconstruction of ultrashort optical pulses, Opt. Lett. 23 (1998) 792–794. https://doi.org/10.1364/OL.23.000792.

[8] J.W. Nicholson, J. Jasapara, W. Rudolph, F.G. Omenetto, A.J. Taylor, Full-field characterization of femtosecond pulses by spectrum and cross-correlation measurements, Opt. Lett. 24 (1999) 1774. https://doi.org/10.1364/OL.24.001774.

[9] A. Gliserin, S.H. Chew, S. Kim, D.E. Kim, Complete characterization of ultrafast optical fields by phase-preserving nonlinear autocorrelation, Light Sci. Appl. 11 (2022) 277. https://doi.org/10.1038/s41377-022-00978-3.

[10] H.K. Tsang, J.B.D. Soole, L.Y. Chan, H.P. LeBlanc, M.A. Koza, R. Bhat, High sensitivity autocorrelation using two-photon absorption in InGaAsP waveguides, Electron. Lett. 31 (1995) 1773–1775. https://doi.org/10.1049/el:19951185.

[11] J.M. Roth, T.E. Murphy, C. Xu, Ultrasensitive and high-dynamic-range two-photon absorption in a GaAs photomultiplier tube, Opt. Lett. 27 (2002) 2076. https://doi.org/10.1364/OL.27.002076.

[12] W. Rudolph, M. Sheik-Bahae, A. Bernstein, L.F. Lester, Femtosecond autocorrelation measurements based on two-photon photoconductivity in ZnSe, Opt. Lett. 22 (1997) 313. https://doi.org/10.1364/OL.22.000313.

[13] M. Piccardo, N.A. Rubin, L. Meadowcroft, P. Chevalier, H. Yuan, J. Kimchi, F. Capasso, Mid-infrared two-photon absorption in an extended-wavelength InGaAs photodetector, Appl. Phys. Lett. 112 (2018) 041106. https://doi.org/10.1063/1.5018619.

[14] Y. Takagi, S. Imamura, T. Kobayashi, K. Yoshihara, Multiple- and single-shot autocorrelator based on two-photon conductivity in semiconductors, Opt. Lett. 17 (1992) 658. https://doi.org/10.1364/OL.17.000658.

[15] F.R. Laughton, J.H. Marsh, D.A. Barrow, E.L. Portnoi, The two-photon absorption semiconductor waveguide autocorrelator, IEEE J. Quantum Electron. 30 (1994) 838–845. https://doi.org/10.1109/3.286177.

[16] T.K. Liang, H.K. Tsang, I.E. Day, J. Drake, A.P. Knights, M. Asghari, Silicon waveguide two-photon absorption detector at 1.5 μm wavelength for autocorrelation measurements, Appl. Phys. Lett. 81 (2002) 1323–1325. https://doi.org/10.1063/1.1500430.

[17] L.P. Barry, B.C. Thomsen, J.M. Dudley, J.D. Harvey, Autocorrelation and ultrafast optical thresholding at 1.5μm using a commercial InGaAsP 1.3μm laser diode, Electron. Lett. 34 (1998).

[18] J.M. Dudley, D.T. Reid, W. Sibbett, L.P. Barry, B. Thomsen, J.D. Harvey, Commercial Semiconductor Devices for Two Photon Absorption Autocorrelation of Ultrashort Light Pulses, Appl. Opt. 37 (1998) 8142. https://doi.org/10.1364/AO.37.008142.

[19] P.J. Maguire, L.P. Barry, T. Krug, W.H. Guo, J. O'Dowd, M. Lynch, A.L. Bradley, J.F. Donegan, H. Folliot, Optical signal processing via two-photon absorption in a semiconductor microcavity for the next generation of high-speed optical communications network, J. Light. Technol. 24 (2006) 2683–2692. https://doi.org/10.1109/JLT.2006.875202.

[20] S.-I. Shin, Y.-S. Lim, Simple Autocorrelation Measurement by Using a GaP Photoconductive Detector, J. Opt. Soc. Korea 20 (2016) 435–440. https://doi.org/10.3807/JOSK.2016.20.3.435.

[21] R. Salem, T.E. Murphy, Polarization-insensitive cross correlation using two-photon absorption in a silicon photodiode, Opt. Lett. 29 (2004) 1524. https://doi.org/10.1364/OL.29.001524.

[22] A.M. Streltsov, J.K. Ranka, A.L. Gaeta, Femtosecond ultraviolet autocorrelation measurements based on two-

photon conductivity in fused silica, Opt. Lett. 23 (1998) 798. https://doi.org/10.1364/OL.23.000798.
[23] J.D. Harvey, J.M. Dudley, B.C. Thomsen, L.P. Barry, Ultra-sensitive optical autocorrelation using two photon absorption, in: OFCIOOC Tech. Dig. Opt. Fiber Commun. Conf. 1999 Int. Conf. Integr. Opt. Opt. Fiber Commun., IEEE, San Diego, CA, USA, 1999: pp. 2–4. https://doi.org/10.1109/OFC.1999.768045.
[24] T. Krug, M. Lynch, A.L. Bradley, J.F. Donegan, L.P. Barry, H. Folliot, J.S. Roberts, G. Hill, High-Sensitivity Two-Photon Absorption Microcavity Autocorrelator, IEEE Photonics Technol. Lett. 16 (2004) 1543–1545. https://doi.org/10.1109/LPT.2004.827102.
[25] K. Kikuchi, Highly sensitive interferometric autocorrelator in the optical communication band using Si avalanche photodiode as two-photon absorber, in: OFC 98 Opt. Fiber Commun. Conf. Exhib. Tech. Dig. Conf. Ed. 1998 OSA Tech. Dig. Ser. Vol2 IEEE Cat No98CH36177, Opt. Soc. America, San Jose, CA, USA, 1998: pp. 313–314. https://doi.org/10.1109/OFC.1998.657432.
[26] D. Duchesne, L. Razzari, L. Halloran, R. Morandotti, A.J. Spring Thorpe, D.N. Christodoulides, D.J. Moss, Two-photon photodetector in a multiquantum well GaAs laser structure at 155μm, Opt. Express 17 (2009) 5298. https://doi.org/10.1364/OE.17.005298.
[27] J.K. Ranka, A.L. Gaeta, A. Baltuska, M.S. Pshenichnikov, D.A. Wiersma, Autocorrelation measurement of 6-fs pulses based on the two-photon-induced photocurrent in a GaAsP photodiode, Opt. Lett. 22 (1997) 1344. https://doi.org/10.1364/OL.22.001344.
[28] S. Lochbrunner, P. Huppmann, E. Riedle, Crosscorrelation measurements of ultrashort visible pulses: comparison between nonlinear crystals and SiC photodiodes, Opt. Commun. 184 (2000) 321–328. https://doi.org/10.1016/S0030-4018(00)00947-0.
[29] C. Xu, Ultra-sensitive autocorrelation of 1.5pm light with single photon counting silicon avalanche photodiode, (n.d.).
[30] P.M.W. Skovgaard, R.J. Mullane, D.N. Nikogosyan, J.G. McInerney, Two-photon photoconductivity in semiconductor waveguide autocorrelators, Opt. Commun. 153 (1998) 78–82. https://doi.org/10.1016/S0030-4018(98)00228-4.
[31] D.T. Reid, M. Padgett, C. McGowan, W.E. Sleat, W. Sibbett, Light-emitting diodes as measurement devices for femtosecond laser pulses, Opt. Lett. 22 (1997) 233. https://doi.org/10.1364/OL.22.000233.
[32] A.K. Sharma, M. Raghuramaiah, P.A. Naik, P.D. Gupta, Use of commercial grade light emitting diode in auto-correlation measurements of femtosecond and picosecond laser pulses at 1054 nm, Opt. Commun. 246 (2005) 195–204. https://doi.org/10.1016/j.optcom.2004.10.067.
[33] T. Fiehler, C. Saraceno, G. Steinmeyer, U. Wittrock, Pitfall in autocorrelation measurements of laser radiation, Opt. Express 32 (2024) 36811. https://doi.org/10.1364/OE.533567.
[34] W.-Y. Choi, MBE-Grown Long Wavelength InGaAlAs/InP Laser Diodes, (n.d.).
[35] D.C. Hutchings, B.S. Wherrett, Theory of anisotropy of two-photon absorption in zinc-blende semiconductors, Phys. Rev. B 49 (1994) 2418–2426. https://doi.org/10.1103/PhysRevB.49.2418.
[36] M.D. Dvorak, W.A. Schroeder, D.R. Andersen, A.L. Smirl, B.S. Wherrett, Measurement of the anisotropy of two-photon absorption coefficients in zincblende semiconductors, IEEE J. Quantum Electron. 30 (1994) 256–268. https://doi.org/10.1109/3.283768.
[37] S. Buckley, M. Radulaski, J. Petykiewicz, K.G. Lagoudakis, J.-H. Kang, M. Brongersma, K. Biermann, J. Vučković, Second-Harmonic Generation in GaAs Photonic Crystal Cavities in (111)B and (001) Crystal Orientations, ACS Photonics 1 (2014) 516–523. https://doi.org/10.1021/ph500054u.
[38] D.N. Fittinghoff, I.A. Walmsley, J.L. Bowie, J.N. Sweetser, R.T. Jennings, M.A. Krumbügel, K.W. DeLong, R. Trebino, Measurement of the intensity and phase of ultraweak, ultrashort laser pulses, Opt. Lett. 21 (1996) 884. https://doi.org/10.1364/OL.21.000884.
[39] A. Weiner, Effect of group velocity mismatch on the measurement of ultrashort optical pulses via second harmonic generation, IEEE J. Quantum Electron. 19 (1983) 1276–1283. https://doi.org/10.1109/JQE.1983.1072036.
[40] A. Baltuška, Z. Wei, M.S. Pshenichnikov, D.A. Wiersma, R. Szipőcs, All-solid-state cavity-dumped sub-5-fs laser, Appl. Phys. B Lasers Opt. 65 (1997) 175–188. https://doi.org/10.1007/s003400050262.
[41] X. Zeng, J. Lontchi, M. Zhukova, P. Bolt, M. Smor, L. Fourdrinier, G. Li, D. Flandre, High-performance dual-mode ultra-thin broadband CdS/CIGS heterojunction photodetector on steel, Opt. Express 30 (2022) 13875. https://doi.org/10.1364/OE.456352.
[42] M. Razeghi, A. Rogalski, Semiconductor ultraviolet detectors, J. Appl. Phys. 79 (1996) 7433–7473. https://doi.org/10.1063/1.362677.
[43] M.I. Saidaminov, Md.A. Haque, M. Savoie, A.L. Abdelhady, N. Cho, I. Dursun, U. Buttner, E. Alarousu, T. Wu, O.M. Bakr, Perovskite Photodetectors Operating in Both Narrowband and Broadband Regimes, Adv. Mater. 28 (2016) 8144–8149. https://doi.org/10.1002/adma.201601235.
[44] M. Kielar, T. Hamid, M. Wiemer, F. Windels, L. Hirsch, P. Sah, A.K. Pandey, Light Detection in Open-Circuit Voltage Mode of Organic Photodetectors, Adv. Funct. Mater. 30 (2020) 1907964. https://doi.org/10.1002/adfm.201907964.
[45] A. Chlebowski, Data for article: Sensitive and accurate femtosecond pulse characterization via two-photon absorption in Fabry-Pérot laser diodes, (2025). https://doi.org/10.18150/HSXYFD.

# Supplementary material: Sensitive and accurate femtosecond pulse characterization via two-photon absorption in Fabry-Pérot laser diodes

This supplement material discusses in detail the process of establishing uncertainty of the pulse duration measurement process.

Measurement of pulse duration through autocorrelation is an indirect process and therefore the following procedure must be implemented for discovery of its uncertainty:

1. Identify sources of uncertainty (Fig. S1),
2. Characterize these sources by finding their standard deviations,
3. Propagate errors through mathematical analysis used for calculating the pulse duration using Monte Carlo method.

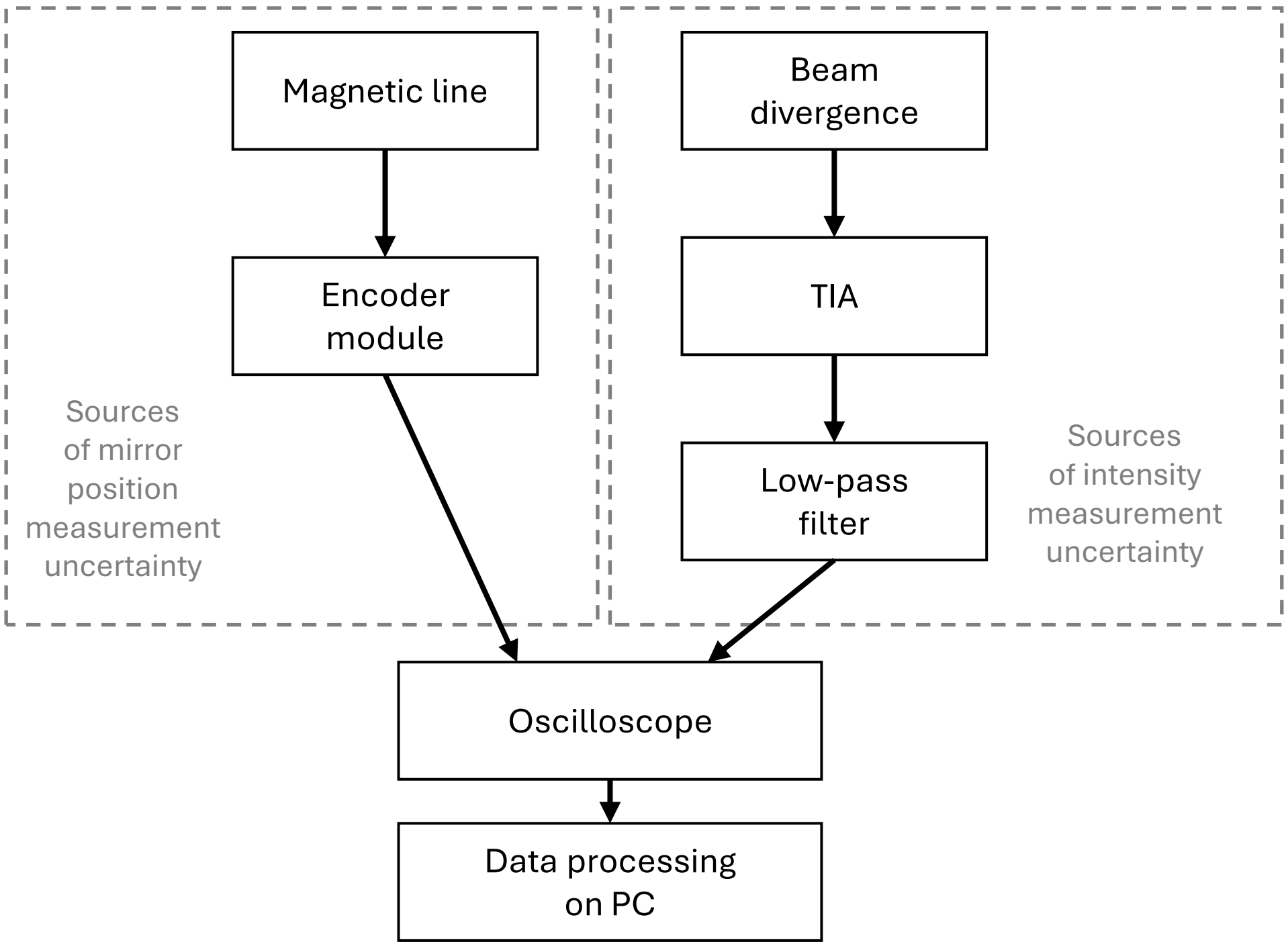

**Fig. S1.** Uncertainty sources

All uncertainty sources presented in Fig. S1 are discussed below.

## 1. Magnetic line and encoder module

The mirror positioning system employed in the autocorrelator is based on a miniature RLB incremental encoder with a magnetic line attached to the delay line's base. Stated relevant parameters of this system are as follows:

| | |
|---|---|
| Base accuracy of the MSD01 magnetic line | ±10 µm / m |
| Hysteresis | ±3 µm |
| Repeatability | ±1 µm |
| External magnetic field tolerance | < 1 mT |
| Magnetic damage threshold | 25 mT |

The total scan length used in the experiment is 4.16 mm, which translates to ±41.6 nm basic precision. Hysteresis does not play a role in the process of acquisition as data is only gathered in scans conducted in one direction. In this situation, repeatability parameter is by far the most significant error source as stated by the documentation.

While this number would normally be sufficient for uncertainty analysis, in this case it was deemed necessary to modify the standard approach and not rely on manufacturer-provided numbers due to a rather low external magnetic field tolerance of less than 1 mT.

The constructed autocorrelator setup is based on a in-house made delay line which employs a voice coil actuator (VCA) with a 60 mm maximum displacement. The electromagnetic half of the VCA is mounted in direct proximity to the magnetic encoder and can therefore influence its operation.

While the VCA documentation does not provide details on its construction, its parameters can be estimated from copper wire resistance and the core diameter with enough precision to judge whether the magnetic field is strong enough to interfere with the positioning system.

Assuming the number of winds N ~ 1000, as approximated using an ohmmeter and calipers, the following calculations can be conducted:

$$B_{\mathrm{max}} = \mu_0 n I_{\mathrm{max}} \, ,$$

where:

$$n = \frac{N}{l} = \frac{1000}{63\ \mathrm{mm}} = 15\,873\ \mathrm{m}^{-1} \, ,$$

$$I_{\mathrm{max}} = 2\ \mathrm{A} \, .$$

Then:

$$B_{\mathrm{max}} = 3.99 \cdot 10^{-2}\ \mathrm{T} \sim 40\ \mathrm{mT} \, .$$

With 40 mT being significantly higher than the tolerance of the encoder, its performance must be verified by other means.

An experiment was conducted where the delay line was integrated into a fiber Mach-Zender interferometer, fed from a stable laser source.

The reference laser source used in the measurement is a 1548.60 nm telecommunications DFB laser diode WDM8 module for Thorlabs PRO 8000 chassis. This laser offers a wavelength stability of ±0.002 nm/24 h, which is insignificant for the purpose of this work and will be disregarded. Accuracy of the wavelength is specified at ±25 pm.

Sinusoidal interferogram (IGM) produced with the reference laser was recorded together with the magnetic encoder signals with a multichannel oscilloscope during a mirror scan conducted in the region where measurements presented in this work were acquired.

The IGM was processed using Hilbert transform and based on its momentary phase and the reference laser wavelength, the true position of the mirror was calculated and compared to the indication of the magnetic encoder.

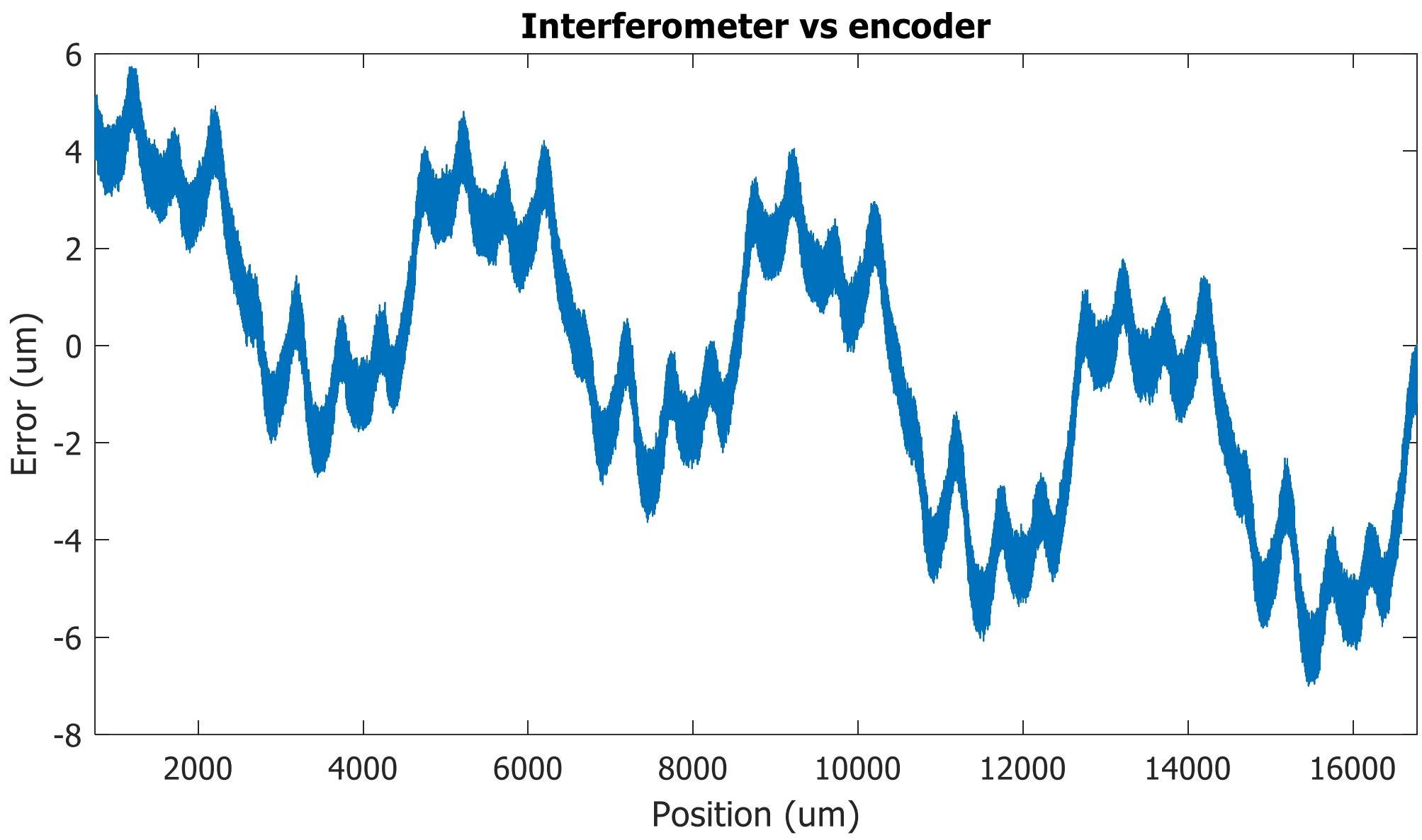

**Fig. S2** Encoder position error.

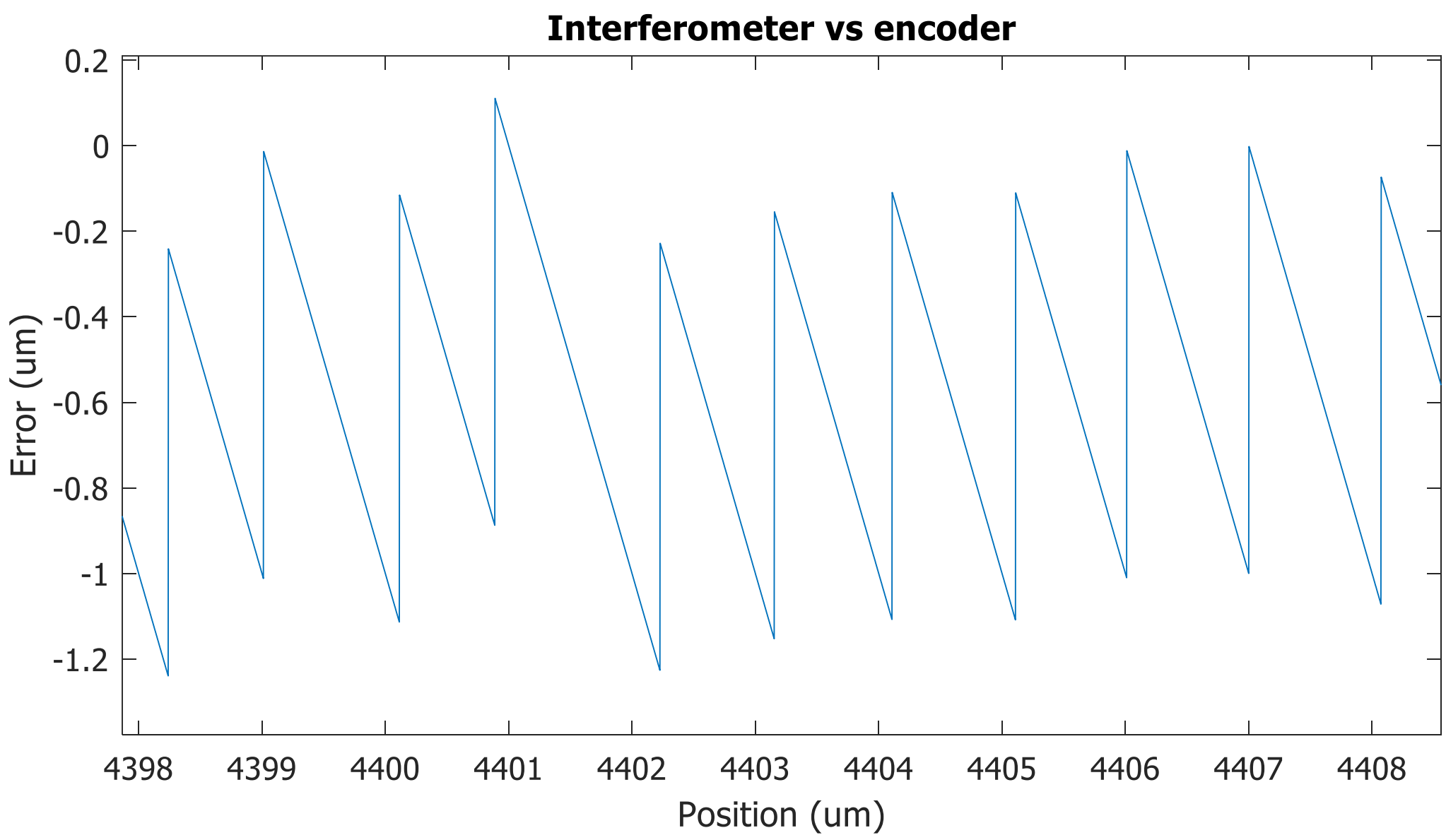

**Fig. S3.** Encoder position error zoomed showing encoder resolution limitation.

The result is the position measurement error curve (Fig. S2-3). The following characteristics can be derived from it:

1. The major, square wave-like periodic error with ~10 µm peak to peak value has a period of 4 mm, which roughly corresponds to the magnetization period of the magnetic line (2 mm per pole).
2. Additional parabolic-shaped errors also seem to be periodic in nature and synchronized to the major error mentioned above. While it has no obvious match to indicate its source this synchronization suggests it likely comes from interpolation performed by the internal logic of the encoder.
3. After sufficient magnification along the x-axis a sawtooth-like, periodic error can be observed. It has exactly 1 µm peak to peak value and 1 µm period. It reflects the 1 µm resolution of the encoder itself, which is much lower than in case of the Hilbert transform-based method, which isn't discrete in nature and relies on momentary IGM phase for position retrieval.
4. The overall error characteristic isn't completely periodic, linear fit reveals a significant drift of 0.415 nm/µm.

After removing the drift, the standard deviation was established to be 2.04 µm.

For purely stochastic, fluctuations classical deviation often fails to converge. This limitation obscures gradual position changes due to drift within the fluctuating signal. To address this, Allan deviation was developed to enable precise measurements by uncovering drift phenomena. It provides a means to extrapolate a system's drift at infinity from finite measurements. Unlike classical deviation, Allan deviation converges to a finite value for common external noise sources, including drift (1/f noise), shot noise, and damped oscillations.

To mitigate the significant statistical error in Allan deviation for large τ, a sliding interval approach can be employed, sacrificing statistical independence for improved accuracy. This method computes the difference between two adjacent intervals, each containing m elements, and then shifts the intervals forward by one element while discarding the oldest data point. By maintaining a constant number of elements per interval, this technique enhances statistical robustness. As a result, the statistical error scales $1/\sqrt{m\,(n-1)}$ instead of the conventional $1/\sqrt{n}$, offering a more reliable estimate for long averaging times.

The conventional approach to standard deviation calculation, defined as the mean over a sliding window of duration τ, is often employed to assess drift in a dataset. However, a fundamental limitation of this standard deviation is its non-convergence for certain types of noise commonly found in nature, such as white noise. Consequently, alternative statistical tools, such as Allan deviation, are necessary to accurately capture low-frequency drift phenomena while ensuring convergence.

The line position values and their associated errors (Fig. S2-3) were translated into optical delay uncertainties (Fig. S4). Analyzing the Allan deviation reveals that increasing the delay (i.e., extending the scan length) directly impacts the measurement error. The minimum observed error corresponds to a ~32 fs pulse duration (7 µm line feed) with a value of ±0.26 fs. However, as the delay increases, the error grows, reaching a maximum of 8.7 fs. Of particular relevance is the delay of 50 fs, at which the primary measurements were conducted. For this value, the Allan deviation is ±0.50 fs, representing a substantial error that cannot be disregarded.

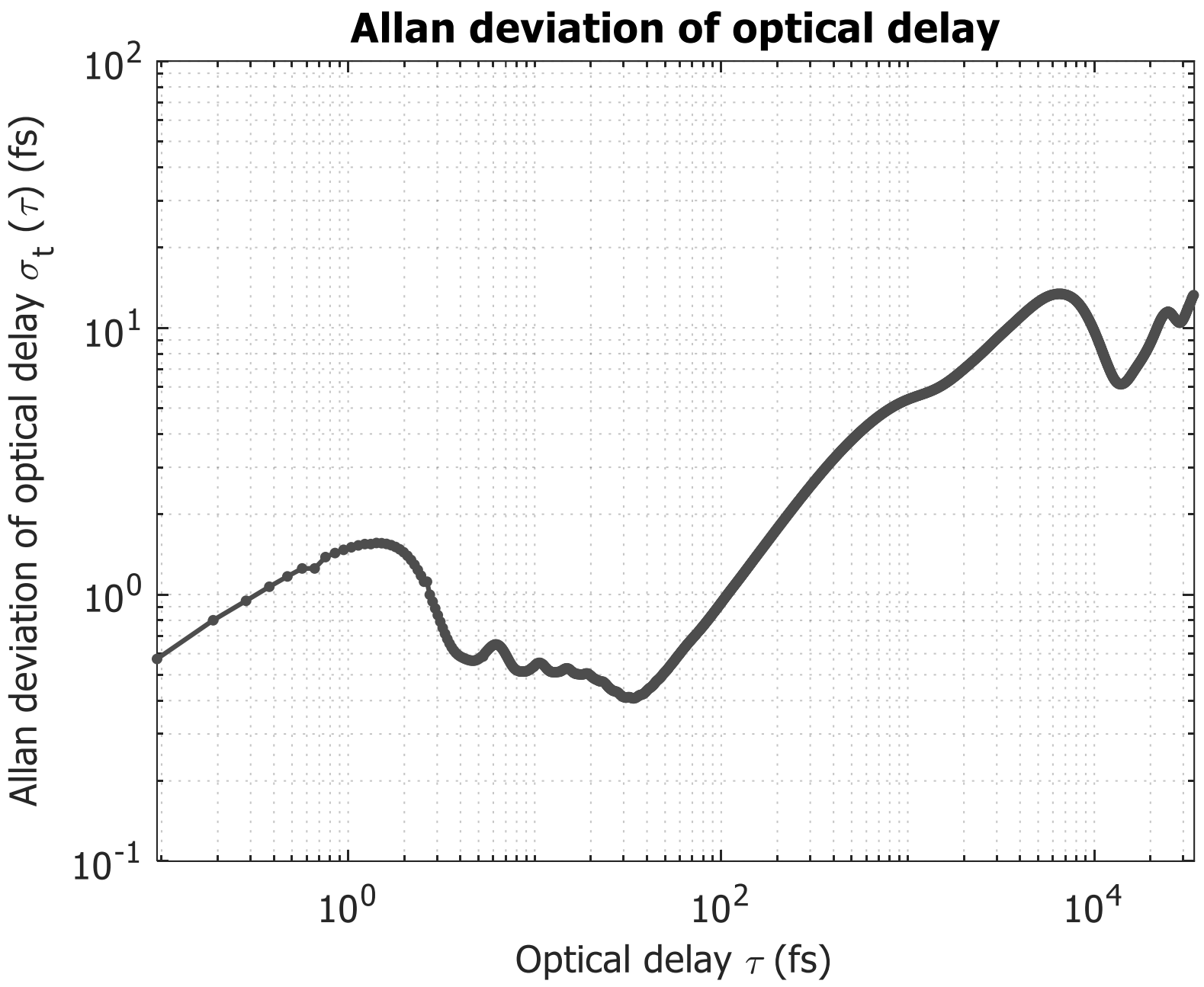

**Fig. S4.** Allan deviation of optical delay.

## 2. Beam divergence

Beam divergence on the other hand is another factor affecting accuracy. Two-photon absorption (TPA) is a strongly intensity-dependent process. Highly divergent beam makes it harder to achieve overlap, leading to weak signals or distorted autocorrelation traces. Moreover increased divergence reduces spatial coherence, leading to imperfect interference patterns in the TPA detector. If divergence causes uneven spatial overlap, the detected pulse duration may appear longer than the actual pulse due to underestimation of peak intensity.

To mitigate the uncertainty as much as possible, a ThorLabs F280APC-1550 fiber collimator was used to ensure collimation before reaching the detector, the beam in front of the detector is focused using a parabolic mirror, and the detector is placed on an XYZ table to ensure optimal distance where beam overlap is maximized.

However, the place that can introduce the highest uncertainty is the delay line and its alignment. To measure its impact, a ThorLabs BP209-IR2/M beam profiler was employed, and placed at the same distance as the detector. The change in centroid X and Y and Peak X and Y were measured for the whole line sweep. Centroid X changed from 522.8 µm to 639.9 µm, with a standard deviation at the 95% confidence level of ±23.9 µm, respectively Centroid Y from 262.8 µm to 368.4 µm with a deviation of ±22.1 µm, Peak X from 618 to 869.4 µm, a deviation of 60.06 µm, Peak Y from 229 µm to 361.8 µm, a deviation of 26.19 µm.

However, these values refer to a full line scan. Variations in the effective range over which the measurement is performed are in the range of ~10 µm with beam diameters X 3089.4 and Y 3616.3 µm. The change in the centroid and peak of the beam over the effective scan range is several orders of magnitude smaller than its diameter. Moreover, considering the beam focus and correction utilizing the XYZ table, the error is many orders of magnitude lower than those generated by the encoder in the X-axis and the oscilloscope at the Y-axis, and therefore may be safely neglected.

## 3. Signal path

The signal path for intensity measurement consists of the detector itself, transimpedance amplifier (TIA) and a low-pass filter (LPF). All these components produce noise which must be accounted for.

Combined spectral noise density of this system was measured at the point of connection to the oscilloscope using a Zurich Instruments MFLI in fast Fourier transform (FFT) mode.

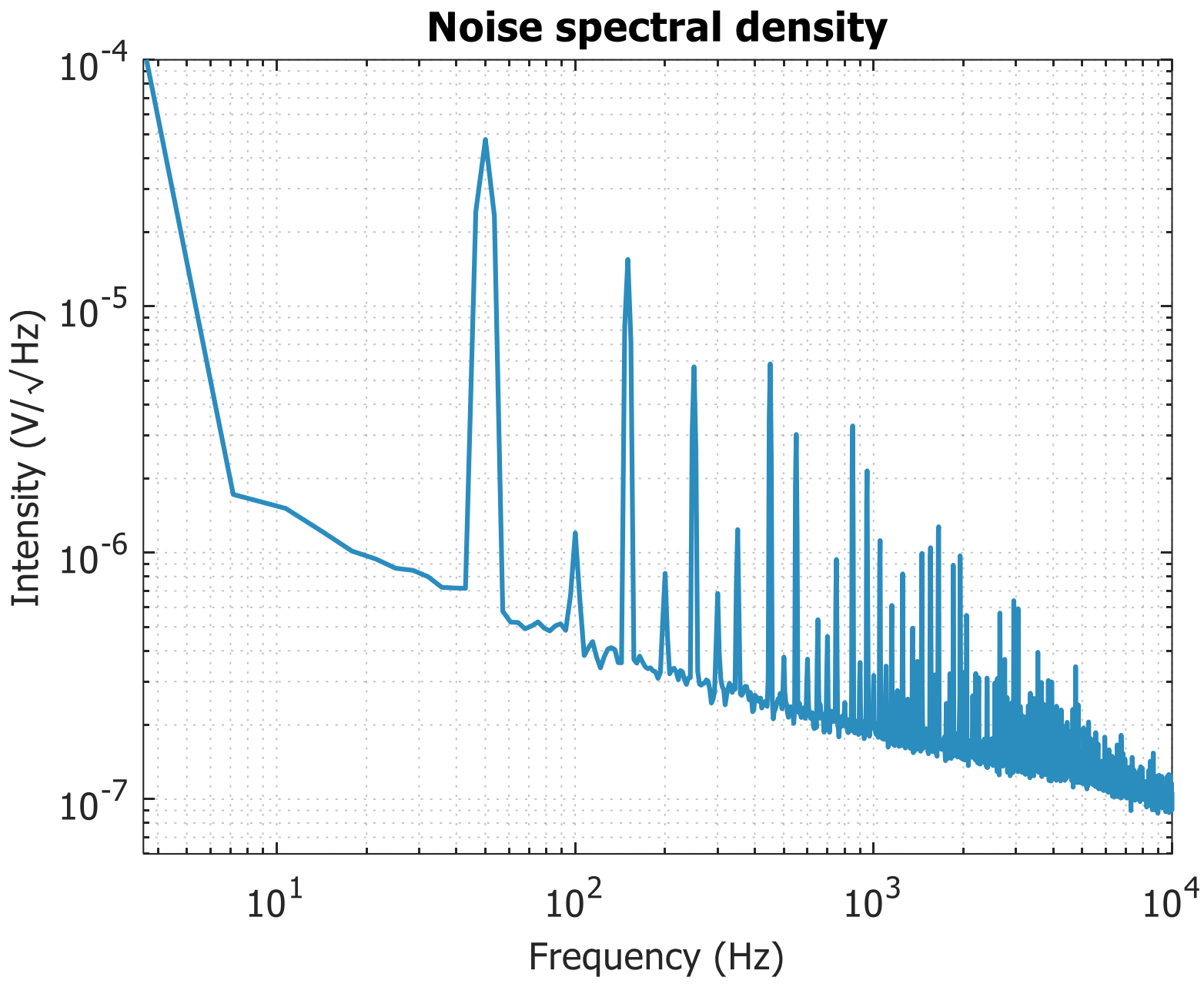

**Fig S5.** Noise spectral density of the signal path of the primary detector

The acquired noise characteristics (Fig. S5) reveal that SNR can be improved by applying improved shielding to the setup, since all captured peaks are likely coupled from external sources. As shown by further analysis, their impact on the quality of the final measurement is far from dominant.

Assuming the noise spectral density is clipped on the side of the higher frequencies by the 3-rd order low pass filter, the overall RMS noise can be calculated as follows:

$$V_{\text{noise RMS}} = \sqrt{\int_{\text{DC}}^{f_{3\text{dB}}} {SD_f}^2 \, \mathrm{d}f}$$

Using numerical integration, the result was found to be: $V_{\text{noise RMS}} = 122\ \mu\text{V}$.

**4. Oscilloscope**

The 4-channel oscilloscope used for data acquisition is Siglent SDS2354X Plus. It acquires signals produced by optoelectronic detectors, as well as A and B signals from the magnetic encoder. It's gain and offset errors do not have an impact on the result of the measurement as establishing the duration of a pulse is not affected by linear errors like this. The oscilloscope introduces two types of random errors which can significantly alter these results through:

- Voltage noise impacts intensity measurement.
- Jitter impacts mirror positioning.

This oscilloscope has an RMS jitter of 10 ps. This is by multiple orders of magnitude less than would be required to match the error produced by the encoder itself and can therefore be safely disregarded.

The voltage noise is significant and was measured for all sensitivities used in the measurement using the standard deviation and statistics measurement built-into the scope.

**Table S1.** Standard deviation of oscilloscope

| Volt/Div | Standard deviation |
|---|---|
| 500 mV | ±4.12 mV |
| 200 mV | ±2.426 mV |
| 100 mV | ±1.501 mV |
| 50 mV | ±0.354 mV |
| 20 mV | ±0.258 mV |
| 10 mV | ±0.221 mV |
| 5 mV | ±0.105 mV |

Subsequently, as in the experimental procedure, the resulted standard deviation is normalized through a scaling ratio to align with the experimental results. The normalized data is then incorporated into the final simulation, ensuring consistency between theoretical modeling and empirical observations.

**5. Data processing on PC**

The complex, nonlinear nature of the algorithms used for calculating the duration of the pulse from raw data makes it impractical to derive the uncertainty in an analytical manner. For this reason, standard deviations obtained from earlier steps are used to produce representative levels of noise, which are then propagated through the algorithm. This approach, based on Monte Carlo method, allows to establish the impact of raw data uncertainty on the final result.

The factor has a definite impact on pulse width and the error introduced by the magnetic line. While the error due to amplitude fluctuations is minimal and does not significantly influence the result, it has nevertheless been accounted for.

Our simulation employs repeated random sampling to estimate system behaviour under uncertainty. Predefined uncertain parameters generate multiple instances, effectively constructing a Monte Carlo-style analysis. Each sampled instance represents a distinct realization of the uncertain system, enabling comprehensive statistical evaluation.

The generated instances undergo the same processing steps as in the actual measurement. Fringe-resolved autocorrelation signals are filtered using a low-pass filter to derive intensity autocorrelation (IAC). The IAC is then fitted with a $sech^2$ function to determine the full width at half maximum, which, after applying the appropriate deconvolution factor, yields the pulse width. This simulation is repeated $N$ times, and the resulting pulse width values are compiled into a histogram. The final standard deviation is derived from this histogram, ensuring a 95% confidence level in the measurement uncertainty.

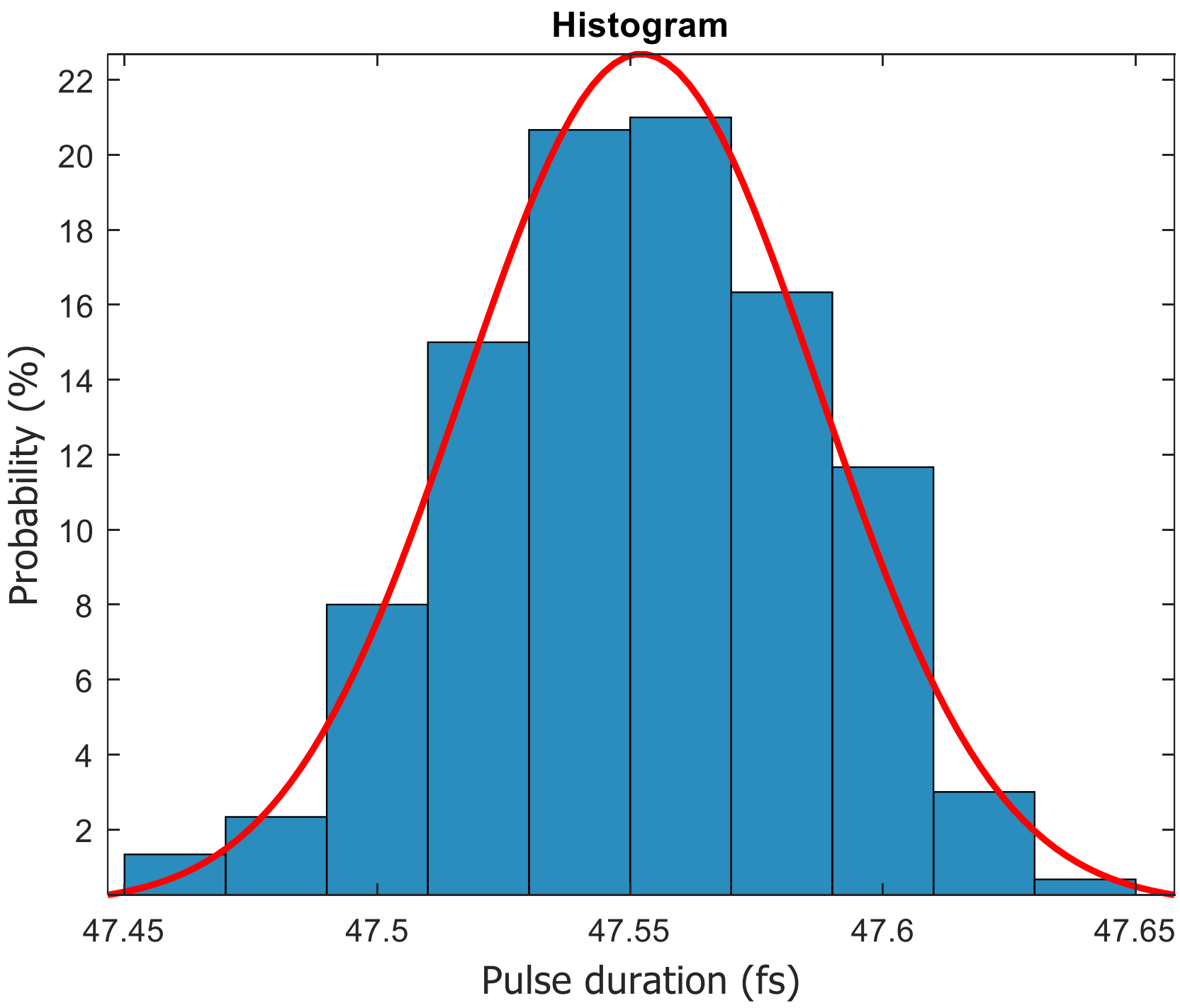

**Fig. S6** Histogram of 300 repeated simulations.

For a sub-47.53 fs pulse duration, 300 simulations were conducted (Fig. S6), yielding a histogram that closely follows a Gaussian distribution. The standard deviation derived from this analysis is approximately 0.035 fs, while the Allan deviation is 0.33 fs, indicating a negligible impact of the algorithms on deviation fluctuations. All subsequent Monte Carlo analyses followed the same methodology, with a final deviation estimates calculated at a 95% confidence level.

**Table S2:** Pulse duration, Allan deviation and standard deviation based on the histogram

| Pulse duration (fs) | Allan deviation (fs) | Final standard deviation (fs) |
|---|---|---|
| 32.69 | ±0.26 | ±0.056 |
| 47.53 | ±0.33 | ±0.070 |
| 51.83 | ±0.34 | ±0.090 |
| 53.66 | ±0.35 | ±0.090 |
| 58.87 | ±0.38 | ±0.106 |
| 69.22 | ±0.44 | ±0.120 |
| 83.94 | ±0.51 | ±0.136 |
| 105.53 | ±0.63 | ±0.198 |
| 140.97 | ±0.83 | ±0.296 |
| 183.56 | ±1.05 | ±0.334 |
| 232.92 | ±1.32 | ±0.446 |